\documentclass[
journal=ancac3, 
manuscript=article]{achemso}

\setkeys{acs}{maxauthors=20}

\usepackage[version=3]{mhchem} 
\usepackage{siunitx}
\usepackage[english=british]{csquotes}
\usepackage[table,xcdraw]{xcolor}

\usepackage{xr}
\usepackage{booktabs}
\usepackage{multirow}
\usepackage{adjustbox}
\usepackage{hyperref}

\makeatletter
\newcommand*{\addFileDependency}[1]{
    \typeout{(#1)}
    \@addtofilelist{#1}
    \IfFileExists{#1}{}{\typeout{No file #1.}}
}
\makeatother

\newcommand*{\myexternaldocument}[1]{%
    \externaldocument{#1}%
        \addFileDependency{#1.tex}%
        \addFileDependency{#1.aux}%
}

\myexternaldocument{main_SI_revised}

\usepackage{afterpage}

\DeclareSIUnit\Molar{\textsc{m}}
\DeclareSIUnit\atm{atm}

\author{Debora Monego}
\affiliation[University of Sydney]
{ARC Centre of Excellence in Exciton Science, School of Chemistry, The University of Sydney, NSW 2006, Australia}
\alsoaffiliation[Sydney Nano]
{The University of Sydney Nano Institute, The University of Sydney, NSW 2006, Australia}

\author{Thomas Kister}
\affiliation[INM --- Leibniz Institute for New Materials]
{INM --- Leibniz Institute for New Materials, Campus D2 2, 66123 Saarbr\"ucken, Germany}

\author{Nicholas Kirkwood}
\affiliation[University of Melbourne]
{ARC Centre of Excellence in Exciton Science, School of Chemistry, University of Melbourne, Parkville, Victoria 3010, Australia}

\author{David Doblas}
\affiliation[INM --- Leibniz Institute for New Materials]
{INM --- Leibniz Institute for New Materials, Campus D2 2, 66123 Saarbr\"ucken, Germany}

\author{Paul Mulvaney}
\affiliation[University of Melbourne]
{ARC Centre of Excellence in Exciton Science, School of Chemistry, University of Melbourne, Parkville, Victoria 3010, Australia}

\author{Tobias Kraus}
\affiliation[INM --- Leibniz Institute for New Materials]
{INM --- Leibniz Institute for New Materials, Campus D2 2, 66123 Saarbr\"ucken, Germany}
\alsoaffiliation[Saarland University]{Colloid and Interface Chemistry, Saarland University, Campus D2 2, 66123 Saarbr\"ucken, Germany}

\author{Asaph Widmer-Cooper}
\email{asaph.widmer-cooper@sydney.edu.au}
\affiliation[University of Sydney]
{ARC Centre of Excellence in Exciton Science, School of Chemistry, The University of Sydney, NSW 2006, Australia}
\alsoaffiliation[Sydney Nano]
{The University of Sydney Nano Institute, The University of Sydney, NSW 2006, Australia}

\newcommand{\myParagraph}{This document is the unedited Author’s version of a Submitted Work that was subsequently accepted for publication in ACS Nano, \copyright  2020 American Chemical Society after peer review. To access the final edited and published work see \href{https://dx.doi.org/10.1021/acsnano.9b03552}{\textcolor{blue}{\textit{ACS Nano} \textbf{2020}, 14, 5278-5287}}.}

\title{When Like Destabilizes Like: Inverted Solvent Effects in Apolar Nanoparticle Dispersions}

\begin{document}

\afterpage{
  \noindent
  \myParagraph
  \vspace*{\fill}
  \thispagestyle{empty} 
  \addtocounter{page}{-1} 
  \newpage
}

\maketitle

\section{Abstract}
We report on the colloidal stability of nanoparticles with alkanethiol shells in apolar solvents. Small angle X-ray scattering and molecular dynamics simulations were used to characterize the interaction between nanoparticles in linear alkane solvents ranging from hexane to hexadecane, including \SI{4}{\nano\meter} gold cores with hexadecanethiol shells and \SI{6}{\nano\meter} cadmium selenide cores with octadecanethiol shells. We find that the agglomeration is enthalpically driven and that, contrary to what one would expect from classical colloid theory, the temperature at which the particles agglomerate increases with increasing solvent chain length. We demonstrate that the inverted trend correlates with the temperatures at which the ligands order in the different solvents, and show that the inversion is due to a combination of enthalpic and entropic effects that enhance the stability of the ordered ligand state as the solvent length increases. We also explain why cyclohexane is a better solvent than hexane, despite having very similar solvation parameters to hexadecane.

\section{Keywords}
nanoparticle, dispersion, apolar, colloidal stability, ligand, solvent, agglomeration

\section{Introduction}

Inorganic nanoparticles made of metals \cite{Jana2003, Stoeva2002}, semiconductors \cite{Peng1998, Shim2000}, and oxides \cite{Pileni2003} are now used as functional components in catalysis \cite{Murray1993, Haruta2003}, sensing \cite{Shipway2000, Saha2012}, photovoltaics \cite{Talapin2005, Zhao2015}, and color conversion in white light generation \cite{Chen2006a, Chen2006b, Achermann2006}. Many applications require the particles to be dispersed individually in organic solvents or to pass through this stage during processing.

Purely inorganic particles do not form stable dispersions in apolar solvents because van der Waals (vdW) forces cause attraction and thus agglomeration of the particles \cite{Korgel1998}. Cores are therefore coated with organic molecules during synthesis \cite{Brust1994, Puntes2001, Murray1993, Peng1998} or during subsequent ligand-exchange procedures. The adsorbed ligands provide steric stabilization and reduce the interfacial energy of the particles \cite{Napper1977}.

The colloidal stability of ligand-coated particles in small molecule solvents is commonly explained with the classical \enquote{like dissolves like} rule, whereby the colloid interaction is assumed to be purely repulsive in solvents that are good for the tail group of the ligands \cite{Sigman2004, Bodnarchuk2010, Schapotschnikow2008}. Reducing the quality of the solvent, in turn, induces enthalpic attraction between the ligands \cite{Patel2007, Khan2009, Dalmaschio2013, Widmer-Cooper2014, Moura2015} and is a common way of destabilizing these suspensions \cite{Talapin2001}. Van der Waals attraction between the cores can also drive agglomeration, even in good solvents, if the cores are sufficiently large or the ligands are too short \cite{ Abecassis2008, Khan2012, Kister2018}. 

Surprisingly, there appear to be exceptions to the rule of \enquote{like dissolves like} even for small metal and semiconductor particles whose agglomeration solely depends on the ligand shell \cite{Kister2018,Monego2018}. Lohman \textit{et al.} found that gold nanoparticles with octane- or hexadecanethiol shells were more stable in alkanes shorter than the ligand chain \cite{Lohman2012}, while Hajiw \textit{et al.} found that gold nanoparticles with hexane- or dodecanethiol shells were more stable in cyclohexane than in heptane or dodecane, respectively \cite{Hajiw2015}. In polymer solutions and melts, where the conformational entropy of free polymer molecules can drive particle agglomeration \cite{deGennes1980,Koski2019}, colloidal stability does typically decrease with polymer length \cite{Sunday2012, Sunday2015}. However, a purely entropic explanation seems unlikely in relatively short solvents, like the ones described above, where the agglomeration is driven by enthalpy.

Here, we have studied the dispersibility of gold and cadmium selenide nanoparticles in a variety of linear and cyclic alkane solvents using experiments that characterize their temperature-dependent colloidal stability. In all cases, colloidal stability decreased as the length of the alkane solvent approached that of the ligand tail, opposite to the rule of \enquote{like dissolves like}. Further, we found that cyclohexane is a considerably better solvent for the particles than hexadecane, despite the two solvents having very similar solvation parameters. Specifically, changing the solvent from hexadecane to cyclohexane decreased the agglomeration temperature by \SI{15}{\celsius}. It is important to understand the origin of such inversions, because the choice of solvents for stable dispersions of nanoparticles is of considerable practical relevance; it affects the quality of nanocomposites, nanocrystal assembly \cite{Zou2017}, and phase transfer procedures \cite{Fini2008}, ultimately affecting device processability and performance.

In order to understand the origin of this behavior, we compared our systematic experimental data with detailed molecular dynamics simulations. Our results indicate that the inversion is a consequence of both enthalpic and entropic effects that together enhance the stability of an attractive ligand state as the solvent chain length increases. As has been shown previously, ligand shells composed of linear alkyl tails can undergo an ordering transition in solution that switches the interaction between the nanoparticles from repulsive to attractive \cite{Widmer-Cooper2014}. The temperature of this ligand phase transition is sensitive to various parameters including the particle dimensions, density of ligand coverage, and ligand length, often leading to non-linear trends that cannot be explained using classical colloid theory \cite{Widmer-Cooper2016, Kister2018, Monego2018}. We show that even small changes in the solvent structure can strongly impact the ligand ordering transition and use this insight to explain how particles can have dramatically different interactions in solvents with almost the same Hamaker, Hildebrand and Hansen parameters. 

\section{Results and Discussion}

\begin{figure}[htpb]
    \centering
    \includegraphics[width=\linewidth]{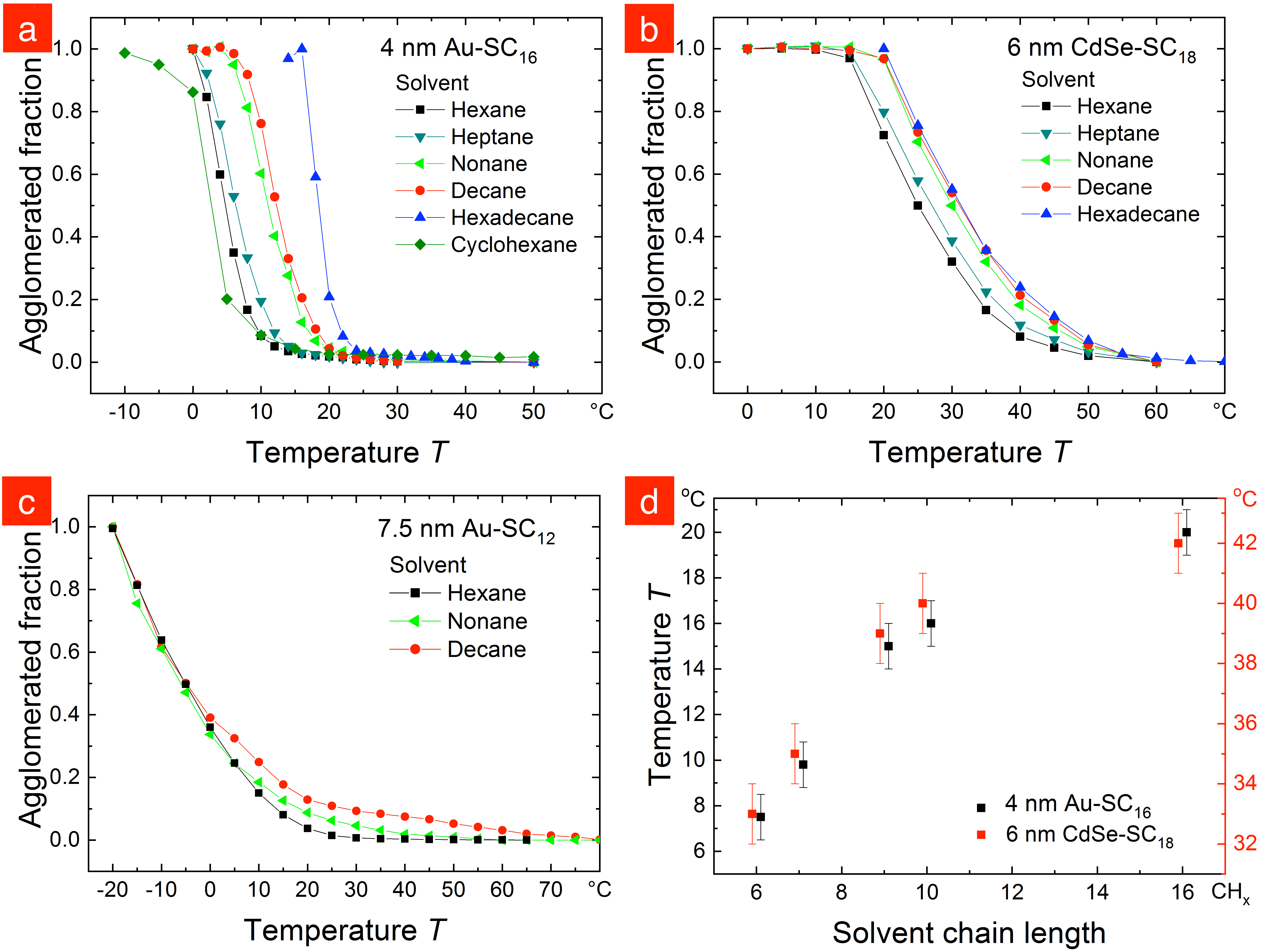}
    \caption{Fraction of agglomerated (a) \SI{4}{\nano\meter} \ce{Au-SC_{16}}, (b) \SI{6}{\nano\meter} \ce{CdSe-SC_{18}}, and (c) \SI{7.5}{\nano\meter} \ce{Au-SC_{12}} particles, as determined by \textit{in situ} small angle X-ray scattering. All particles were dispersed at high temperatures, and agglomeration occurred upon cooling as indicated by the increase in structure factor. (d) Agglomeration temperature (where \SI{20}{\percent} of particles were agglomerated) as a function of alkane solvent chain length.}
    \label{Figure_1}
    \vspace{0.5em}
\end{figure}

Nanoparticles (NP) with \ce{Au} cores (\SI{4}{\nano\meter} and \SI{7.5}{\nano\meter} in diameter) and \ce{CdSe} cores (\SI{6}{\nano\meter} in diameter) were coated with hexadecanethiol (\ce{SC_{16}}), dodecanethiol (\ce{SC_{12}}), and octadecanethiol (\ce{SC_{18}}) chains, respectively. These particles were dispersed in linear and cyclic alkane solvents of different chain lengths and analyzed by \textit{in situ} small angle X-ray scattering (SAXS) at a concentration of \SI{2.5}{\milli\gram\per\milli\liter} (roughly \SI[mode=text]{3.8E15}{NP \milli\litre^{-1}}). For all solvents tested, the particles agglomerated below a certain temperature (Figure \ref{Figure_1}), which was observed as a peak in the structure factor\cite{Johnson1959} \(S(q)\). The agglomeration temperature, \(T_{agglo}\), defined as the temperature at which \SI{20}{\percent} of the particles had agglomerated, increased in all cases as the solvent length approached the ligand length (Figure \ref{Figure_1}d). Similar results were obtained for \SI{7.5}{\nano\meter} \ce{Au} cores coated with hexadecanethiol (\ce{SC_{16}}) ligands (see Figure S\ref{SI_7p5nmSC16} in the Supporting Information). No sign of solvent freezing was observed in any of the experiments.

These results show that the agglomeration is enthalpically favorable and entropically unfavorable, with cooling required to destabilize the dispersions. Extensions of classical colloid theory to describe the colloidal stability of ligand-coated nanoparticles in solution, for example reference \citenum{Khan2009}, typically include two terms that favor dispersion (the ideal entropy of mixing and the conformational entropy of the ligands) and two terms that favor agglomeration (the vdW attraction between the cores and the non-ideal free energy of mixing). The vdW attraction between the cores is insignificant for our particles,\cite{Kister2018} leaving the free energy of mixing as the deciding term in this classical approach. This can be described by the Flory-Huggins theory (equations \ref{equation:free_energy_of_mixing1} and \ref{equation:free_energy_of_mixing2}) that quantifies the affinity of the tethered ligands for the solvent in terms of the ideal entropy of mixing and the Flory $\chi$ parameter (equation \ref{equation: flory_parameter}). In these equations, \(d\) is the core diameter, \(\tilde{L}\) is the rescaled ligand length (ligand length divided by core diameter), \(k_b\) the Boltzmann's constant, \(T\) the absolute temperature, \(\phi\ = \left( N_L \frac{\nu_L}{V_{Sh}} \right)^2\) is the volume fraction occupied by the ligand shell, \(\nu_S\) is the volume of a solvent molecule, \(\nu_L\) the volume of a ligand molecule, \(N_L\) the number of ligands per nanoparticle, \(V_{Sh}\) the volume of the ligand shell, \(V_S\) is the molar volume of the solvent, \(R\) is the universal gas constant, and \(\delta_L\) and \(\delta_S\) are the Hildebrand solubility parameters for the ligands and solvent, respectively.

\begin{equation}
    \frac{G_{mix}}{k_bT}= \frac{\pi d^3}{2\nu_S} \phi^2 \left( \frac{1}{2}- \chi \right) \left( \tilde{s}- 1 -2\tilde{L} \right)^2; \quad 1+\tilde{L}<x<1+2\tilde{L}
    \label{equation:free_energy_of_mixing1}
\end{equation}
\begin{equation}
    \frac{G_{mix2}}{k_bT}= \frac{\pi d^3}{\nu_S} \phi^2 \tilde{L}^2 \left( \frac{1}{2}- \chi \right) \left( 3ln \frac{\tilde{L}}{\tilde{s}-1} + 2\frac{\tilde{s}-1}{\tilde{L}} - \frac{3}{2} \right); \quad x<1+\tilde{L}
    \label{equation:free_energy_of_mixing2}
\end{equation}
\begin{equation}
    \chi = \frac{V_S}{RT} (\delta_L- \delta_S)^2 + 0.34
    \label{equation: flory_parameter}
\end{equation}

A Flory parameter below \(0.5\) indicates that the free energy of mixing of solvent and ligand is negative and that the two components should spontaneously mix. Since only the alkane tails of the ligands interact with the solvent, it seems reasonable to approximate the solubility parameters of the ligands by those of the unthiolated alkanes, \textit{i.e.} hexadecane (\SI{16.4}{{\mega\pascal}^{1/2}}) and octadecane (\SI{17.1}{{\mega\pascal}^{1/2}}). Thus, hexadecane is expected to be a better solvent for these coatings than decane (\(\delta_{s} = \SI{15.8}{{\mega\pascal}^{1/2}}\)) and hexane (\(\delta_{s} = \SI{14.9}{{\mega\pascal}^{1/2}}\)) (Hildebrand parameters from reference \citenum{Brandrup1999}), even taking into account the reduction in the ideal entropy of mixing ($\Delta S_{mix}^{ideal}$) as the solvent chain length increases (see Table S\ref{SI_Table_SolvParams} in the Supporting Information). A similar conclusion is reached when considering Hamaker constants or Hansen solubility parameters of the ligand and solvent molecules. Even if one were to use different solubility parameters for alkyl ligands bound to nanoparticles, it would remain impossible within this theoretical framework to explain why cyclohexane is a much better solvent than hexadecane, since the two have almost identical solubility parameters.

An alternative explanation for deviations from the rule of \enquote{like dissolves like} in short-chain solvents was proposed by Hajiw and co-workers, who studied the temperature-dependent dispersibility of small Au particles (roughly \SI{2.3}{\nano\meter} in diameter) in a similar range of alkane solvents.\cite{Hajiw2015} They noted that the colloidal stability of polymer-grafted particles in polymer melts does typically decrease with polymer length and speculated that similar thermodynamic driving forces may explain such trends in much shorter solvents. In polymer melts, however, it is the conformational entropy of the free polymer chains that drives agglomeration. As we shall show, the solvent conformational entropy ($ S_{conf}^{solv}$) in short-chain solvents is much smaller and unable to explain the observed dispersibility trends. In order to quantify $S_{conf}^{solv}$, along with enthalpic effects that are not considered in classical colloid theory, we used molecular dynamics (MD) simulations (see Methods for details).

\begin{figure}[htpb]
    \centering
    \includegraphics[width=\linewidth]{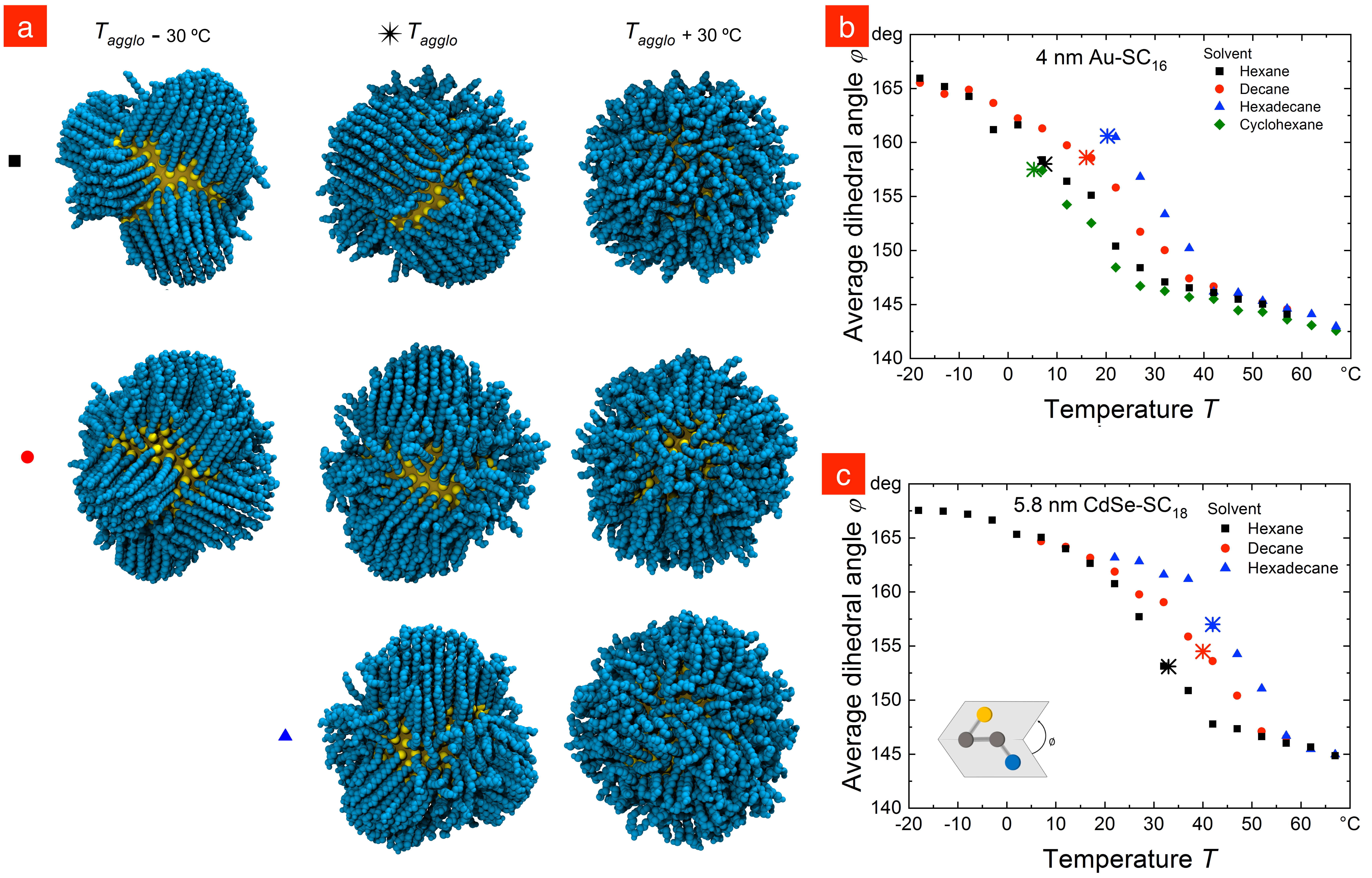}
    \caption{(a) Simulation snapshots of \SI{4}{\nano\meter} Au particles at \(T_{agglo}\) \(\pm\) \SI{30}{\celsius} in hexane, decane, and hexadecane. Solvent molecules have been hidden for clarity, with symbols as shown in the plot legends. The ligands order as the temperature decreases in a similar way in all solvents. This transition can be quantified by the average dihedral angle of the ligands, which increases rapidly as they order for both (b) \SI{4}{\nano\meter} \ce{Au-SC_{16}} and (c) \SI{5.8}{\nano\meter} \ce{CdSe-SC_{18}} particles. For comparison, the experimental agglomeration temperatures have been indicated by large crossed symbols. The scheme at the bottom left of (c) shows the definition of the dihedral angle \(\phi\).}
    \label{Figure_2}
    \vspace{0.5em}
\end{figure}

MD simulations of \SI{4}{\nano\meter} core diameter \ce{Au} and \SI{5.8}{\nano\meter} \ce{CdSe} particles in explicit solvent show that upon cooling, the ligands adopt more extended configurations and cluster together into ordered bundles in an enthalpically driven process. Snapshots of the simulations above, at, and below \(T_{agglo}\) in the linear alkane solvents are shown in Figure \ref{Figure_2}a and in Figure S\ref{SI_CdSe_snapshots} in the Supporting Information, with solvent molecules hidden for clarity. Similar ligand shell structures were found for cyclohexane (shown in Figure S\ref{SI_Cyclohexane} in the Supporting Information). The average dihedral angle of the ligand tails was used to quantify the ordering transition in the shell, with the results shown in Figures \ref{Figure_2}b and \ref{Figure_2}c. The simulations show that the same ligand shells \enquote{order} at higher temperature in longer alkane solvents for both \ce{Au} and \ce{CdSe} cores. This trend mirrors the experimental results, with the experimentally observed particle agglomeration temperatures (indicated by large crossed symbols) always occurring after the ligands have started ordering. These results indicate that particle agglomeration is driven by the ligand shell transition regardless of the core material, solvent length or structure, consistent with our previous findings for similar particles in decane \cite{Kister2018} and with earlier experimental results for much larger silica particles in hexadecane \cite{Roke2006}.

\begin{figure}[htpb]
    \centering
    \includegraphics[width=0.6\linewidth]{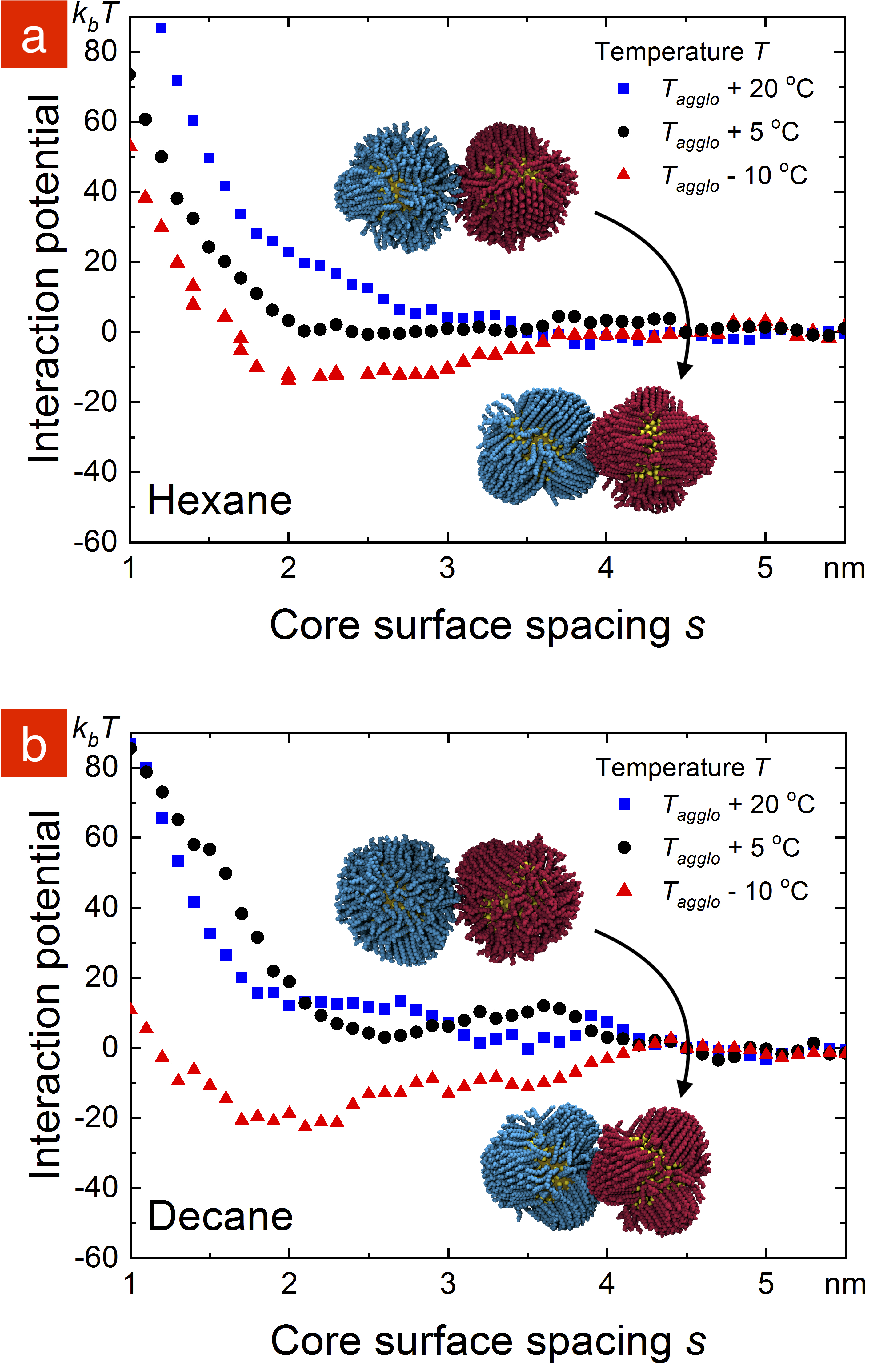}
    \caption{Potentials of mean force calculated for pairs of \SI{4}{\nano\meter} \ce{Au-SC_{16}} particles in (a) hexane and (b) decane, at temperatures around \(T_{agglo}\). Simulation snapshots show the state of the ligands above and below \(T_{agglo}\), at temperatures corresponding to the blue squares and red triangles, respectively.}
    \label{Figure_3}
    \vspace{0.5em}
\end{figure}

To explicitly test whether ordering of the ligand shell drives agglomeration regardless of solvent length, we calculated the potential of mean force between pairs of \SI{4}{\nano\meter} \ce{Au} particles in explicit hexane and decane (Figures \ref{Figure_3}a and \ref{Figure_3}b, respectively) as a function of separation and temperature. The overall interaction switched from repulsive to attractive as the ligands ordered, irrespective of solvent. This change in interaction between the ligand shells arises from changes in how the ligands interact with one another and with the solvent as their conformational state changes (see Figure S\ref{SI_PMFComponents} in the Supporting Information). The ligand-ligand component (which includes vdW interactions between the ligands) becomes more attractive as the ligands order, while the ligand-solvent component (which includes entropy changes involving the solvent) becomes less repulsive. In contrast, the overall interaction between alkyl ligand shells is always attractive in the absence of solvent, regardless of their conformational state, due to the absence of competing ligand-solvent interactions \cite{Schapotschnikow2008,Waltmann2018,Widmer-Cooper2014,Kister2018}. (The relevant results in references \citenum{Widmer-Cooper2014} and \citenum{Kister2018} are located in the Supporting Information of those papers.)

\begin{figure}[htpb]
    \centering
    \includegraphics[width=\linewidth]{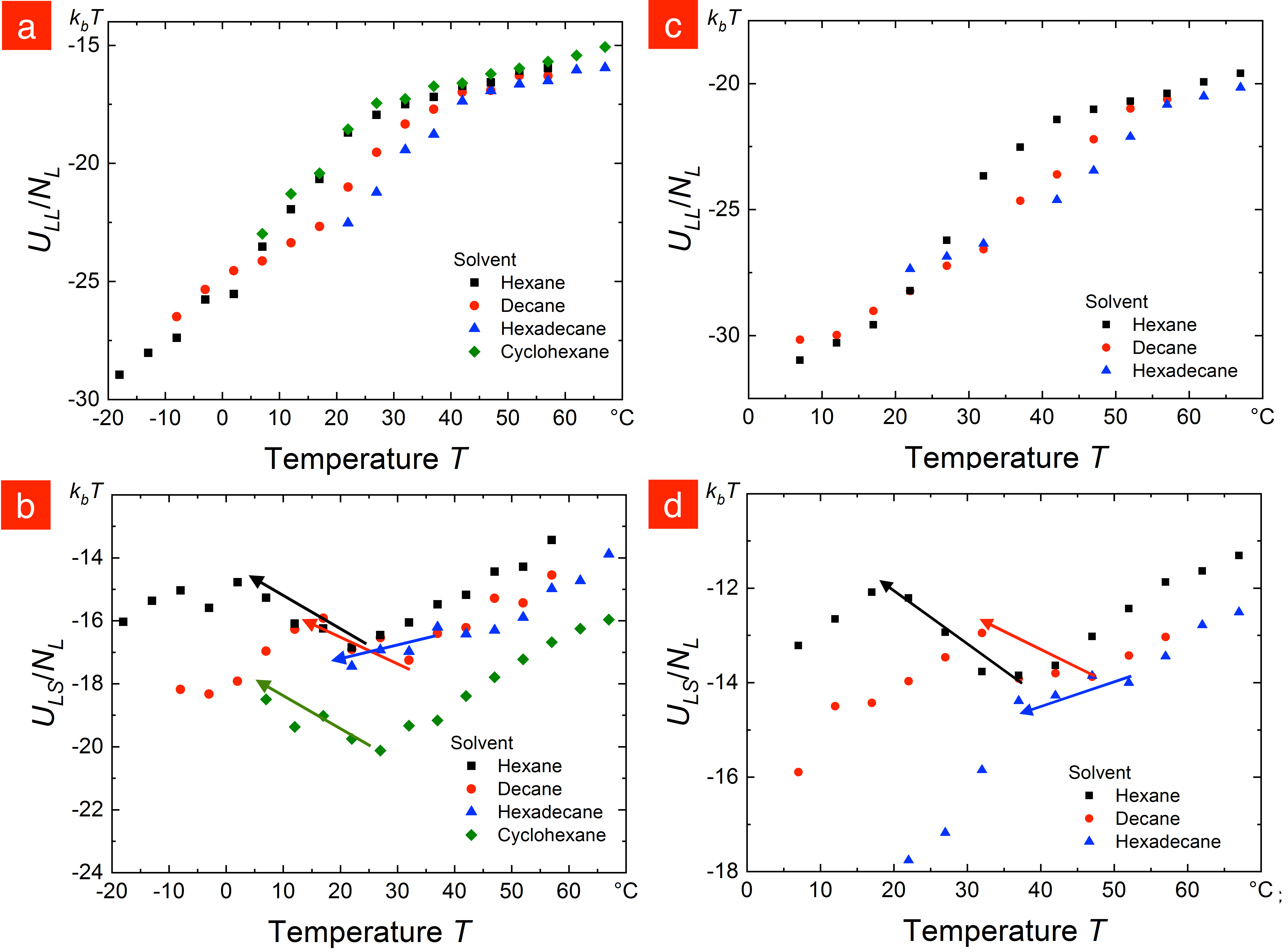}
    \caption{The energy of interaction between ligand molecules (\(U_{LL}\)) on isolated (a) \SI{4}{\nano\meter} \ce{Au-SC_{16}} and (c) \SI{5.8}{\nano\meter} \ce{CdSe-SC_{18}} particles increases upon cooling in all solvents. In contrast, the energy of interaction between ligand and solvent molecules (\(U_{LS}\)) decreases in magnitude for shorter chain alkanes and increases for hexadecane for both (b) \ce{Au} and (d) \ce{CdSe} nanoparticles, as indicated by the solid arrows. All energies are normalized by the number of ligand molecules on the nanoparticle, \(N_{L}\).}
    \label{Figure_4}
    \vspace{0.5em}
\end{figure}

So far, we have established that the nanoparticles agglomerate because the ligands order and that the enthalpic driving force for both of these processes is the attractive vdW force that exists between the bound ligands. We now focus on the thermodynamics of the ligand ordering transition in order to explain the origin of the inverted trend in agglomeration temperatures. In particular, we will show that the trend is a consequence of both enthalpic and entropic effects that together enhance the stability of the attractive ligand state as the solvent chain length increases. We start by considering a single nanoparticle and quantifying the dominant enthalpic contributions to the free energy difference between the disordered and ordered ligand states, \textit{i.e.} the vdW interactions between ligand molecules within the same shell (\(U_{LL}\)) and between ligand and solvent molecules (\(U_{LS}\)). These quantities are compared as a function of temperature and solvent type in Figure \ref{Figure_4}.

We find that the interaction between the ligands follows a similar trend in all solvents apart from a temperature offset: \(U_{LL}\) increases in magnitude in all cases by \SI{3}{}-\SI[mode=text]{4}{\(k_bT\)\per ligand} during the transition (Figures \ref{Figure_4}a and \ref{Figure_4}c). Statistical analysis of the interatomic spacings within the ligand shell (see Figure S\ref{SI_fll_frequency} in the Supporting Information) indicate that the structure of the ligand shell is almost identical regardless of solvent when the temperature is expressed relative to \(T_{order}\), defined as the temperature at which the average dihedral angle equals \ang{155} (which corresponds roughly to the middle of the transition). 

In contrast, we find more substantial differences in the interaction between ligand and solvent molecules. Figures \ref{Figure_4}b and \ref{Figure_4}d show that the ligand-solvent interaction energy (\(U_{LS}\)) increases during the ordering transition in hexadecane (blue arrows) but decreases in decane, hexane and cyclohexane (red, black and green arrows), with the largest decrease in hexane (around \SI[mode=text]{2}{\(k_bT\)\per ligand}). This means that the change in internal energy driving the ligand shell to order is reduced in shorter solvents, which partly explains why the particles are more stable in hexane and cyclohexane than in hexadecane.

The differences in \(\Delta U_{LS}\) during the transition are due to subtle molecular effects. As the ligands order to form dense bundles, the solvent molecules within the spherical shell occupied by the ligands become confined to the spaces between the bundles. This results in a similar increase in the radially averaged solvent density near the core surface in all solvents (Figures S\ref{SI_Densities} and S\ref{SI_Cyclohexane} in the Supporting Information), but with very different consequences depending on how well the solvent molecules pack with the ligand bundles. Hexane and cyclohexane pack less well with the ordered ligands, resulting in a loss of ligand-solvent interaction, while hexadecane packs better with the ordered ligands, resulting in a gain in ligand-solvent interaction (Figures S\ref{SI_fls_distribution} and S\ref{SI_Cyclohexane} in the Supporting Information). 

\begin{figure}[htpb]
    \centering
    \includegraphics[width=0.6\linewidth]{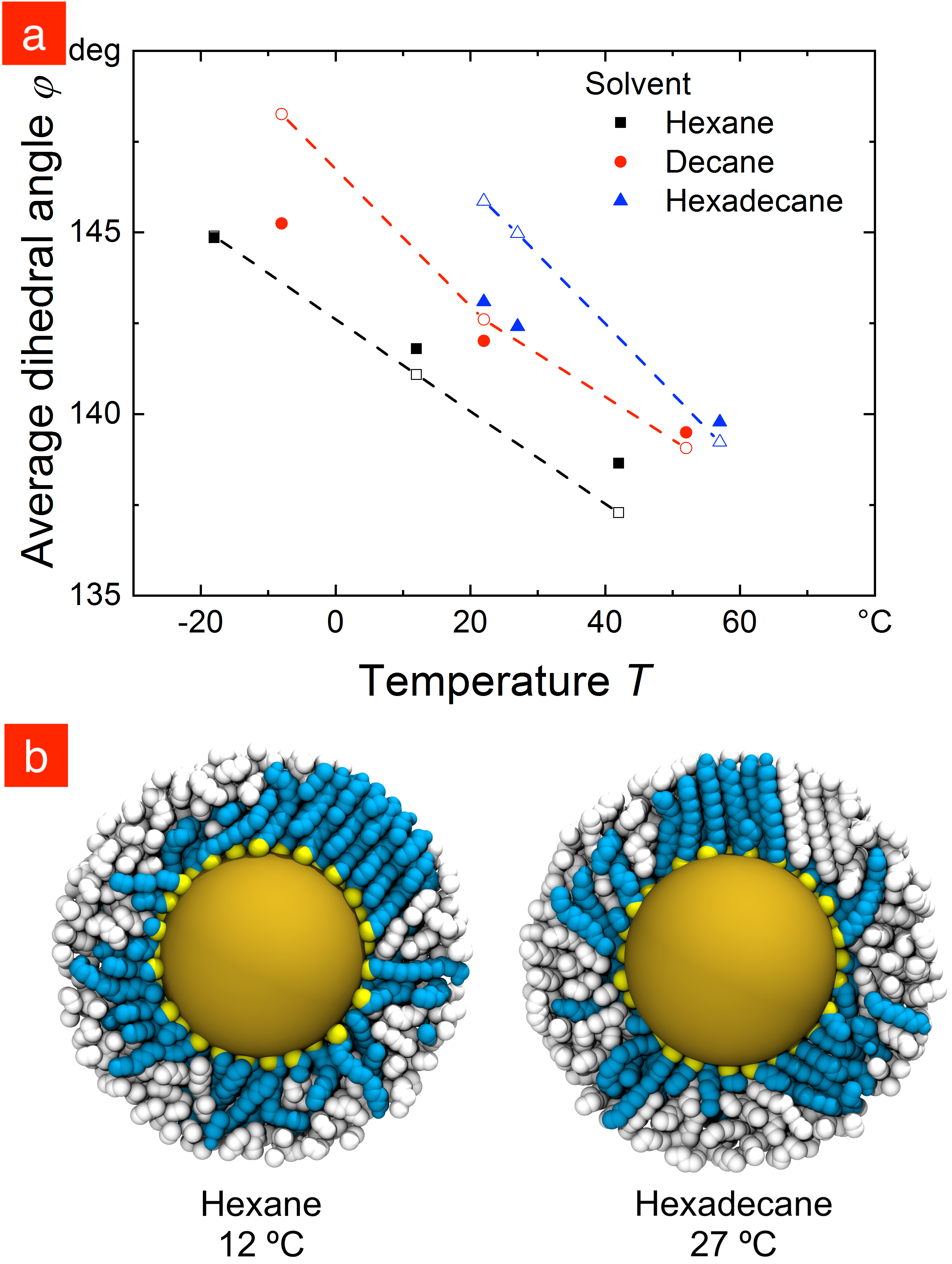}
    \caption{(a) The average dihedral angle for the solvent chains around \ce{Au}NP shows that longer alkanes are more extended at a given temperature, both within the ligand shell (open symbols) and in the bulk solvent region (closed symbols). This allows hexadecane to better align with and stabilize the ligands in the ordered state, as seen in (b) snapshots of the ligand-solvent packing at \(T_{order}\) for \SI{4}{\nano\meter} \ce{Au-SC_{16}} particles. Ligand and solvent united-atoms are represented by blue and white spheres, respectively. The error bars in (a) are smaller than the symbols, and the lines are a guide to the eye.}
    \label{Figure_5}
    \vspace{0.5em}
\end{figure}

Entropic differences between the solvents appear to play an important role here. It is well known that the lower freezing points of shorter chain alkanes are partly due to their higher translational entropy per atom.\cite{Costa2018} Analysis of the average dihedral angles also reveals that longer alkanes are more extended both within and outside the ligand shell, as shown in Figure \ref{Figure_5}a for the \SI{4}{\nano\meter} \ce{Au} NPs. Similar results are obtained for the \ce{CdSe} particles (see Figure S\ref{SI_CdSeDihedral} in the Supporting Information).  These factors may explain why hexadecane is better able than hexane to align with the ordered ligands (Figure \ref{Figure_5}b) and thus increase the relative stability of the ordered state. We note that close alignment of hexadecane with linear alkyl ligands has also been observed in sum frequency generation spectroscopy studies of silica nanoparticles \cite{Roke2006}. 

To more directly address how entropy affects the ligand ordering transition, we quantified the difference (between the ordered and disordered ligand states) in the ideal entropy of mixing (\(\Delta S_{mix}^{ideal}\)) and in the conformational entropies of the ligands (\(\Delta S_{conf}^{lig}\)) and solvent (\(\Delta S_{conf}^{solv}\)). \(\Delta S_{mix}^{ideal}\) is relevant because ordering of the ligands causes them to demix from the solvent. The change in entropy due to this demixing was estimated using equations \ref{equation:free_energy_of_mixing1} and \ref{equation:free_energy_of_mixing2}, setting \(\chi = 0\) and multiplying by 5 for reasons explained in the Methods. This yielded values for \(-T\Delta S_{mix}^{ideal}\) ranging from roughly 0.3 \(k_bT\)/ligand for hexane to 0.1 \(k_bT\)/ligand for hexadecane, indicating a decreasing penalty for ordering as the solvent length increases.

The changes in the conformational entropies were estimated using an information theoretic approach that is described in detail in the Methods. Both ordered and disordered configurations were generated at the same temperature, \(T_{order}\), in order to exclude contributions due to temperature differences. This was achieved by scaling the interaction energies between non-bonded ligand atoms by \(\pm\) 5\%, which resulted in ligand configurations similar to those at \(T_{order} \mp\) \SI{30}{\kelvin}. This yielded values for \(-T\Delta S_{conf}^{lig}\) ranging from \(2.2-3\) \(k_bT\)/ligand, with no apparent trend with solvent length. This substantial penalty is the main reason why ordering of the ligands has to be enthalpically driven. In comparison, we obtained values for \(-T\Delta S_{conf}^{solv}\) ranging from \(0-0.3\) \(k_bT\)/ligand. Again there was no apparent trend with solvent length, indicating that the conformational entropy of the solvent makes at best a small contribution to the trend in agglomeration temperatures.

To facilitate comparison, we estimated the equivalent differences in \(U_{LL}\) and \(U_{LS}\) at \(T_{order}\) \textit{via} linear extrapolation of the data points obtained above and below the ordering transition, and collected all of the enthalpic and entropic terms in Table \ref{tab:deltaG}. Also included is the substantial change in the internal energy associated with the ligand dihedral angles (\(\Delta U_{dih}^{lig}\)), obtained from the same biased simulations as the conformational entropies. Together, these values indicate that the main contribution to the inverted trend in the agglomeration temperatures is the reduction in \(\Delta U_{LS}\) as the solvent length increases, with a minor contribution from the reduction in \(-T\Delta S_{mix}^{ideal}\).

\begin{table}[]
\centering
\caption{Major enthalpic and entropic contributions to the difference in free energy between the ordered and disordered ligand states, expressed in units of \(k_bT\)/ligand at the ligand ordering temperature $T_{order}$: $U_{LL}$ is the vdW interaction between the ligands, $U_{LS}$ is the vdW interaction between the ligands and the solvent, $U_{dih}^{lig}$ is the internal energy due to the ligand dihedral angles, $S_{mix}^{ideal}$ is the entropy of demixing the ligands and solvent, while $S_{conf}^{lig}$ and $S_{conf}^{solv}$ are the conformational entropies of the ligands and solvent, respectively. Negative quantities favor the ordered state while positive ones favor the disordered one. The quantities highlighted in green are the only ones that explain the trend in the agglomeration temperatures. The * indicates a value that was estimated by comparison with the same results obtained for CdSe. Standard errors are $\pm0.3$ or smaller.}
\label{tab:deltaG}
\begin{tabular}{|l|r|r|r|}
\hline
\multicolumn{4}{|c|}{\textbf{\SI{4}{\nano\meter} \ce{Au}-\ce{SC16}}} \\ \hline
Solvent & \multicolumn{1}{l|}{Hexane} & \multicolumn{1}{l|}{Decane} & \multicolumn{1}{l|}{Hexadecane} \\ \hline
\(T_{order}\) & 290 & 295 & 300 \\ \hline
\(\Delta U_{LL}\) & -3.8 & -3.7 & -3.5 \\ \hline
\rowcolor[HTML]{32CB00} 
\(\Delta U_{LS}\) & 4.0 & 2.3 & 1.0\(^*\) \\ \hline
\(\Delta U_{dih}^{lig}\) & -2.6 & -2.3 & -1.8 \\ \hline
\rowcolor[HTML]{32CB00} 
\(-T\Delta S_{mix}\) & 0.32 & 0.22 & 0.14 \\ \hline
\(-T\Delta S_{conf}^{lig}\) & 2.5 & 3.0 & 2.5 \\ \hline
\(-T\Delta S_{conf}^{solv}\) & 0.00 & 0.00 & 0.30 \\ \hline
\hline \hline \hline
\multicolumn{4}{|c|}{\textbf{\SI{5.8}{\nano\meter} \ce{CdSe}-\ce{SC18}}} \\ \hline
Solvent & \multicolumn{1}{l|}{Hexane} & \multicolumn{1}{l|}{Decane} & \multicolumn{1}{l|}{Hexadecane} \\ \hline
\(T_{order}\) & 300 & 310 & 320 \\ \hline
\(\Delta U_{LL}\) & -4.4 & -3.5 & -3.1 \\ \hline
\rowcolor[HTML]{32CB00} 
\(\Delta U_{LS}\) & 4.3 & 2.6 & 1.3 \\ \hline
\(\Delta U_{dih}^{lig}\) & -1.9 & -1.8 & -1.3 \\ \hline
\rowcolor[HTML]{32CB00} 
\(-T\Delta S_{mix}\) & 0.54 & 0.37 & 0.25 \\ \hline
\(-T\Delta S_{conf}^{lig}\) & 2.2 & 2.8 & 2.2 \\ \hline
\(-T\Delta S_{conf}^{solv}\) & -0.17 & 0.12 & -0.09 \\ \hline
\end{tabular}
\end{table}

Our results also explain why alkanethiol-coated nanoparticles are more stable in cyclohexane than in hexadecane, despite the two solvents having almost identical density and solubility parameters. In cyclohexane, the reduction in \(U_{LS}\) upon ordering is greater than in hexadecane (due to poorer packing with the extended ligands), while the entropic cost of demixing is also greater due to cyclohexane's smaller size and thus higher number density. The ordered ligand state is therefore less stable (relative to the disordered one) in cyclohexane than in hexadecane, which suppresses the ordering transition and results in an agglomeration temperature that is even lower than that in hexane.

Finally, to check whether \(S_{conf}^{solv}\) contributes to the trend in agglomeration temperatures regardless of the conformational state of the ligands, we calculated the difference in \(S_{conf}^{solv}\) between the solvated \SI{4}{\nano\meter} \ce{Au} nanoparticles and pure solvent with the same number of solvent molecules. This yielded a similar value in all solvents for \(-T\Delta S_{conf}^{solv}\) of roughly 0.2 \(k_bT\)/ligand when the ligands were in the disordered state, indicating no substantial contribution to the trend.

\section{Conclusions}

We have studied the temperature-dependent agglomeration of small apolar metallic and semiconducting nanoparticles in a range of alkane solvents. We found that the agglomeration is enthalpically driven, with colloidal stabilities that run counter to expectations from classical colloid theory. Increasing the solvent chain length towards that of the ligands resulted in a \emph{decrease} in colloidal stability, rather than the expected increase, and the colloidal stability differed strongly between cyclohexane and hexadecane despite their almost identical solvation parameters. 

While this behavior is reminiscent of colloidal stability in polymer solutions and melts, the thermodynamic origins are different, with enthalpic rather than entropic effects dominating in small organic solvents. Simulations show that the nanoparticles become attractive to one another as the ligands order and that the temperatures at which the particles agglomerate match the temperatures at which the ligands order in the various solvents. This indicates that the experimental results can be understood by considering the thermodynamics of the ordering transition and, in particular, how well the various solvents stabilize the ordered state of the ligands relative to the disordered one.

We found that the ordering transition is driven by vdW attraction between the ligand tails, and internal energy associated with their dihedral angles, and opposed by a combination of enthalpic and entropic terms: loss of vdW interaction between the ligands and solvent, loss of ligand conformational entropy, and a reduction in the entropy of mixing. Of these, the loss in vdW interaction with the solvent exhibits the biggest differences, with smaller losses observed as the solvent length increases due to better packing with the ordered ligands. The entropic cost of demixing the ligands from the solvent also exhibits a small decrease as the size of the solvent molecules increases. Together, these changes increase the relative stability of the ordered ligand state when the solvent is changed from hexane or cyclohexane to hexadecane, which explains the experimental results.

We hope that these results will inspire more detailed experimental studies of ligand morphology and ligand-solvent interactions, which are now possible due to recent advances in Sum Frequency Generation spectroscopy \cite{Roke2006}, Nuclear Magnetic Resonance \cite{Roo2018, Ruks2019}, and Small Angle Neutron Scattering \cite{Luo2018}. While we have not considered polar solvents in the present study, we note that linear \ce{-(CH_{2})_{n}X} ligands with a variety of terminal \ce{X} groups can also order in water \cite{Lane2010,Bolintineanu2014}, raising the possibility of similar effects in polar solvents.

\section{Methods}

All chemicals were obtained from Sigma Aldrich (unless noted otherwise) and used without further purification. The methods were chosen to provides samples that are as comparable as possible to data published previously \cite{Kister2018,Monego2018}.

\subsection{Nanoparticle synthesis} 

The synthesis of gold nanoparticles (AuNP) with core diameters of \SI{4}{\nano\meter} and \SI{7.5}{\nano\meter} was adapted from a previously described method \cite{Kister2018}. For AuNPs with a core diameter of \SI{4}{\nano\meter} a mixture of pentane ($\ge$ 98.5\%, GC), \SI{8}{\milli\liter} oleylamine (technical grade, 70\%), and \SI{100}{\milli\gram} of \ce{HAuCl4} (with crystal water) was stirred at \SI{20}{\celsius} and \SI{500}{\radian\per\minute} for \SI{10}{\minute} under argon atmosphere. Afterwards a solution of \SI{40}{\milli\gram} tert-butylamine borane (ABCR, 97\%) in \SI{2}{\milli\liter} pentane and \SI{2}{\milli\liter} oleylamine was added. The color of the solution immediately changed. After stirring for \SI{60}{\minute} at \SI{20}{\celsius}, the nanoparticles were purified once by precipitating with \SI{30}{\milli\liter} ethanol and centrifugation at \SI{4000}{\radian\per\minute} for \SI{5}{\minute}. The precipitated nanoparticles were then redispersed in \SI{20}{\milli\liter} of the appropriate solvent. Gold cores with a diameter of \SI{7.5}{\nano\meter} were produced in benzene instead of pentane. The mixture of benzene, oleylamine and \ce{HAuCl4} was stirred for \SI{1}{\minute} before tert-butylamine borane was added. The resulting dispersion was then purified as above.

Cadmium selenide nanoparticles (CdSeNPs) with core diameters of \SI{6}{\nano\meter} were synthesized as follows. First, three stock solutions were prepared, a \ce{Se} injection solution (i), a \ce{Cd} growth solution (ii), and a \ce{Se} growth solution (iii): 
(i) \SI{0.3265}{\gram} Se were dissolved in a mixture of \SI{2.5}{\gram} trioctylphosphine, \SI{2.5}{\gram} octadecene, and \SI{6}{\gram} oleylamine in a nitrogen-filled glovebox to give a clear, slightly yellow solution. (ii) A solution containing \SI{0.17}{\mol\per\liter} of cadmium was made from \SI{0.22}{\gram} cadmium oxide, \SI{0.97}{\gram} oleic acid, and \SI{6.23}{\gram} 1-octadecene in a 3 neck round bottom flask on a Schlenk line. The solution was degassed under vacuum (\(<\) \SI{1}{\milli\bar}) for \SI{60}{\minute} at \SI{80}{\celsius}, heated to \SI{250}{\celsius} and held until clear, then cooled to room temperature. Whilst cooling \SI{1.13}{\milli\liter} of oleylamine were added. The final solution was clear and slightly yellow. (iii) A solution containing \SI{1.7}{\mol\per\liter} of selenium was prepared by dissolving \SI{0.25}{\gram} of selenium in \SI{1.55}{\gram} trioctylphosphine in a nitrogen-filled glovebox to give a clear colourless solution.

The synthesis started with \SI{0.22}{\gram} cadmium oxide, \SI{3}{\gram} oleic acid, and \SI{30}{\gram} octadecene in a 3 neck round bottom flask that was degassed under vacuum (\(<\) \SI{1}{\milli\bar}) for \SI{60}{\minute} at \SI{80}{\celsius}. The mixture was then heated to \SI{260}{\celsius} until a clear solution (iv) had formed. The selenium injection solution (i) was loaded into a \SI{24}{\milli\liter} disposable syringe equipped with a 16 G needle and rapidly injected into the cadmium solution (iv) at \SI{260}{\celsius}. The temperature of the reaction solution was allowed to recover to \SI{250}{\celsius} where it was held for NP growth. After \SI{20}{\minute}, \SI{2}{\milli\liter} of \SI{0.17}{\mol\per\liter} cadmium growth stock (ii) and \SI{0.2}{\milli\liter} of \SI{1.7}{\mol\per\liter} selenium growth stock (iii) were added dropwise to the reaction. The addition of cadmium and selenium growth solutions (ii, iii) was continued every \SI{10}{\minute}. After 3 additions, the reaction was left for a further \SI{10}{\minute} at \SI{250}{\celsius}, then cooled to room temperature. The NPs were washed three times \textit{via} precipitation with acetone and resuspended in toluene.

\subsection{Nanoparticle characterization}

Small Angle X-ray Scattering (Xenocs Xeuss 2.0) and Transmission Electron Microscopy (JEOL JEM 2010) were used to measure the core size of the NPs as previously described\cite{Kister2018}. Scattering data from SAXS was analyzed using SASfit (Version 0.94.6, Paul Scherrer Institute) and TEM micrographs were analyzed using ImageJ distributed by NIH (Version 1.45s).

\begin{table}[ht]
\centering
\small
  \caption{NPs used for this study, with diameters obtained from transmission electron microscopy and small angle X-ray scattering.}
  \label{table_size}
  \begin{tabular*}{0.8\textwidth}{@{\extracolsep{\fill}}cccc}
    \hline
    Number & d (TEM) & d (SAXS)\\
    \hline
    Au01 & 4.1 nm $\pm$ \SI{10.0}{\percent} & 4.1 nm $\pm$ \SI{9.3}{\percent}\\
    Au02 & 7.4 nm $\pm$ \SI{7.4}{\percent} & 7.5 nm $\pm$ \SI{6.7}{\percent}\\
    CdSe & 5.8 nm $\pm$ \SI{7.1}{\percent} & 6.0 nm $\pm$ \SI{9.6}{\percent}\\
    \hline
  \end{tabular*}
\end{table}

\subsection{Ligand exchange}

\textbf{AuNPs.} Ligand exchange on AuNPs was performed as described previously \cite{Kister2016pressure}. AuNPs coated with oleylamine were heated to \SI{80}{\celsius} and an excess of required alkanethiol was added. After stirring for further \SI{10}{\minute}, the particles were purified and redispersed in the appropriate solvent. 

\textbf{CdSeNPs.} As-synthesized CdSeNPs were precipitated with acetone/ethanol and resuspended in a solution of the respective alkanethiol ligand (40 wt-\(\%\) in chloroform) with triethylamine (1 molar equivalent with respect to thiol). The resulting NP dispersion was heated for 3 hours at \SI{45}{\celsius} while stirring. The NPs were then washed \textit{via} precipitation with antisolvent and centrifugation (3,300 x g for \SI{3}{\minute}). The antisolvent was chosen to optimally dissolve excess ligand: 1:1 (v/v) methanol/ethanol mixture for hexanethiol and octanethiol ligands, or 1:1 (v/v) acetone/ethanol mixture for dodecanethiol and longer ligands. The NPs were resuspended again in a solution of ligand (40 wt-\(\%\) in chloroform), stirred at \SI{45}{\celsius} for 2 hours, then washed as before and resuspended in a 0.1 M solution of ligand in chloroform. After stirring at room temperature for 24-48 hours the NPs were washed three times and resuspended in pure chloroform. Chambrier \textit{et al.} have shown this procedure leads to almost complete displacement ($>$ 92\%) of amines by the alkane thiol ligands \cite{Chambrier2015}.

\subsection{Small Angle X-ray Scattering}

Experiments were performed under vacuum using a Xeuss 2.0 from Xenocs SA (Grenoble, France) equipped with a copper $K_\alpha$ X-ray source and a PILATUS 1M detector from DECTRIS (Baden, Switzerland) using a sample-to-detector distance of \SI{1235}{\milli\meter}. 

To prevent solvent evaporation during the measurements, the samples (usually a quantity of \SI{20}{\micro\litre} to \SI{40}{\micro\litre}) were filled into thin-wall glass capillaries (diameter of \SI{2}{\milli\meter}) and sealed with epoxy resin. 

For each measurement, the samples were introduced into a temperature controlled sample holder (Omega CN8200), Peltier-controlled with a temperature range between \SI{-20}{\celsius} and \SI{120}{\celsius}. The measurements started at high temperature to ensure a fully deagglomerated state. Afterwards the temperature was first decreased and later increased in \SI{5}{\celsius} steps. At each step, the samples were first equilibrated (\SI{20}{\minute}) followed by an exposition of \SI{10}{\minute}. 
Data treatment was carried out as described elsewhere \cite{Schnablegger2013saxs, Kister2018}.

\subsection{Molecular dynamics simulations}

Molecular dynamics (MD) simulations with periodic boundary conditions were used to study \SI{4}{\nano\meter} Au nanoparticles and \SI{5.8}{\nano\meter} \ce{CdSe} nanoparticles covered in 1-hexadecanethiol and 1-octadecanethiol ligands, respectively, in the presence of a variety of liquid alkanes (\textit{n}-hexane, \textit{n}-decane, \textit{n}-hexadecane and cyclohexane). The sulfur atoms from the ligands were randomly placed on a spherical shell around the implicit core (\SI{0.15}{\nano\meter} further out) and allowed to find their optimal positions on this shell while subject to a Coulombic interaction with relative permittivity \(\epsilon\) = \SI[mode=text]{10}{} and the RATTLE constraint\cite{Andersen1983}. This produced a shell with approximately equidistant binding sites, with the sulfur atoms subsequently treated as part of the rigid core of the particle. The ligands were irreversibly bound to the \ce{Au} and \ce{CdSe} cores at a surface coverage of \SI[mode=text]{5.5}{ligands \nano\meter^{-2}}, consistent with thermogravimetric analyses of the experimental samples\cite{Monego2018}. \ce{CH_x} groups from ligand and solvent molecules were treated as united atoms and interacted with one another according to the 12-6 Lennard-Jones (LJ) potential and with the implicit cores through a 9-3 LJ potential, as previously employed and described for similar particles\cite{Widmer-Cooper2014,Pool2007,Monego2018}. Additionally, bond stretching, bond bending, and dihedral torsion terms were considered within each molecule, with parameters taken from the TraPPE force field family\cite{Martin1998}.

Simulations of systems with up to 165,000 particles were performed using the LAMMPS molecular dynamics simulation package\cite{Plimpton1995}, at temperatures ranging from roughly the freezing point of the respective solvent to values sufficiently high to have the ligands in the disordered state (up to \SI{340}{\kelvin}). Periodic simulation cells containing individual nanoparticles and the alkane solvent were slowly compressed until the solvent density far from the NP was equal to the experimental density of the pure solvent at the chosen temperature. A preliminary run was performed at constant volume in order to accommodate the particles properly in the simulation cell. The systems were then equilibrated at constant pressure (\SI{80}{\atm}) and temperature, maintained with a Nos\'e-Hoover thermostat and barostat, for at least \SI{12}{\nano\second}. Finally, the relevant data were accumulated and averaged over production runs of \SI{1}{\nano\second}. Average bulk solvent densities for these runs stayed within \SI{1}{\percent} of experimental values for linear alkanes, and \SI{5}{\percent} for cyclohexane. Molecular graphics were produced using Visual Molecular Dynamics (VMD)\cite{Humphrey1996}.

\subsubsection{Potentials of mean force}
The change in free energy as a pair of \SI{4}{\nano\meter} \ce{Au} particles was brought together was calculated in both hexane and decane as a potential of mean force (PMF) using constrained MD. Starting from a non-interacting separation, the particles were brought together at a rate of \SI{1}{\angstrom\per\nano\second}. The particles were allowed to rotate about their centers of mass at each separation \(r\), and subsequent simulations of \SI{10}{\nano\second} or more were performed in order to adequately sample the PMF. Longer runs were necessary particularly at and below \(T_{order}\), where the ligands were less mobile. Additionally, in order to allow the ligands to reorganize and find more stable configurations at these temperatures, we included a thermal annealing step at separations where the ligand shells overlapped. This was done by increasing the temperature of these systems by \SI{50}{\kelvin} over \SI{1}{\nano\second} and subsequently cooling it back to the initial temperature over the course of \SI{3}{\nano\second}.

The spherical gold cores were assumed to interact with each other \textit{via} the Hamaker potential \cite{Hamaker1937}, with a Hamaker constant of \SI{2}{\electronvolt} \cite{Ederth2001}. This approach treats the solvent and ligands as a single continuum, with the interaction constant scaled to include the effect of the hydrocarbon medium.

The PMF between two nanoparticles is given by
\begin{equation}
    \phi_{MF}(r)=\int_{r}^{\infty} F_{mean}(s) ds
\end{equation}
Where \(F_{mean}\) is the average force in the direction of the line connecting the two particles and is given by
\begin{equation}
    F_{mean}(r)=\frac{1}{2}\langle(\Vec{F}_{2}-\Vec{F}_{1})\cdot\Vec{r}\rangle_{NVT}
\end{equation}
In the above, \(\Vec{F}_1\) and \(\Vec{F}_2\) are the total forces acting on the first and second NP, respectively, \(\Vec{r}\) is the unit vector pointing from one particle's center to the other's, and the angular brackets denote an average in the canonical ensemble.
\subsubsection{Entropy of mixing}
The change in the ideal entropy of mixing due to spatially separating the ligand and solvent molecules was estimated using equations \ref{equation:free_energy_of_mixing1} and \ref{equation:free_energy_of_mixing2} by setting \(\chi = 0\) and calculating the value at the average core spacing in the experimental agglomerates (\SI{2.4}{\nano\meter}). This value was then multiplied by 5 to roughly convert the entropy change for a pair of touching particles into the entropy change for an entire particle capable of accommodating approximately 10 nearest neighbors. The other parameters used are listed in Table S\ref{tab:mixing_parameters} in the Supporting Information.
\subsubsection{Conformational entropies}
We employed the correlation corrected multibody local approximation (CC-MLA), as implemented in the software CENCALC \cite{Suarez2013}, to estimate changes in the conformational entropies of the ligand and solvent molecules. The molecular conformational space was represented in terms of the dihedral angles, which were discretized into subintervals delimiting the three locally stable conformational states accessible to them, \textit{i.e.} \textit{trans}, \textit{gauche}(-), and \textit{gauche}(+). This transforms the continuous random variable \(\theta\) into the discrete random variable \(X\), with a probability mass function \(P(X)\), where \(X = \left\{ X_1, ..., X_M \right\}\) and \(M\) is the number of dihedral angles. The entropy can then in principle be calculated as a sum over the \(N\) possible configurations of the system using the expression for the Shannon information entropy:

\begin{equation}
    \centering
    S_{conform} = -k_b \sum^{N} P \left( X \right) \ln{P \left( X \right)}.
    \label{eq:shannon_entropy}
\end{equation}

In practice, obtaining an accurate estimate for \(P(X)\) is difficult, because the number of possible conformers is very large (\(\sim 3^M\)). Direct application of the Shannon expression would also result in large and negatively-biased entropies due to correlations between the dihedral angles. However, an approximation to the total entropy can be obtained by truncating a mutual information expansion (MIE)\cite{Suarez2011}, allowing the Shannon information entropy to be calculated using a reasonable number of states.

In addition, due to the large number of molecules in our system, each molecule was treated as an independent system; \textit{i.e.}, a nanoparticle coated with \(l\) ligands, each with \(d\) dihedral angles, was analysed as \(l\) independent \(d\)-dimensional spaces. The sum of the entropies of the individual ligands then provided an approximation for the entropy of the entire nanoparticle. This approach considered the correlations within each molecule, but ignored correlations between neighbor molecules, which are stronger for ligands in the ordered state. The conformational entropy differences that we report for the ligands therefore represent a lower bound, calculated using fully converged values from data sampled every \SI{1}{\pico\s} over \SI{5.5}{\nano\s}. While we were not able to fully converge the absolute solvent entropies using the same number of data points, the entropy differences converge more rapidly and do appear to be fully converged.

\acknowledgement
P.M., N.K., D.M. and A.W. were supported by the ARC Centre of Excellence in Exciton Science (CE170100026). A.W. thanks the Australian Research Council for a Future Fellowship (FT140101061), and D.M. thanks the University of Sydney Nano Institute for a Postgraduate Top-Up Scholarship and the Australian Nanotechnology Network for an Overseas Travel Fellowship. Computational resources were generously provided by the University of Sydney HPC service, the National Computational Infrastructure National Facility (NCI-NF) Flagship program, and the Pawsey Supercomputer Centre Energy and Resources Merit Allocation Scheme. T.K., D.D. and T.K. thank the DFG Deutsche Forschungsgemeinschaft for funding. P.M. and T.K. also thank the DAAD for travel support.

\section*{Supporting Information Available}

Supporting Information shows SAXS data, fraction of agglomerated \SI{7.5}{\nano\meter} \ce{Au-SC_{16}} particles, simulation results for \ce{CdSe-SC_{18}} particles dispersed in linear alkanes and for \ce{Au-SC_{16}} particles in cyclohexane, radial density distributions of ligand and solvent, individual contributions to the the PMF (in hexane), and number of ligand-ligand interactions and ligand-solvent interaction energy as a function of the separation between interacting pairs. This material is available free of charge \textit{via} the Internet at http://pubs.acs.org/.

\bibliography{bib}

\providecommand{\latin}[1]{#1}
\makeatletter
\providecommand{\doi}
  {\begingroup\let\do\@makeother\dospecials
  \catcode`\{=1 \catcode`\}=2 \doi@aux}
\providecommand{\doi@aux}[1]{\endgroup\texttt{#1}}
\makeatother
\providecommand*\mcitethebibliography{\thebibliography}
\csname @ifundefined\endcsname{endmcitethebibliography}
  {\let\endmcitethebibliography\endthebibliography}{}
\begin{mcitethebibliography}{2}
\providecommand*\natexlab[1]{#1}
\providecommand*\mciteSetBstSublistMode[1]{}
\providecommand*\mciteSetBstMaxWidthForm[2]{}
\providecommand*\mciteBstWouldAddEndPuncttrue
  {\def\EndOfBibitem{\unskip.}}
\providecommand*\mciteBstWouldAddEndPunctfalse
  {\let\EndOfBibitem\relax}
\providecommand*\mciteSetBstMidEndSepPunct[3]{}
\providecommand*\mciteSetBstSublistLabelBeginEnd[3]{}
\providecommand*\EndOfBibitem{}
\mciteSetBstSublistMode{f}
\mciteSetBstMaxWidthForm{subitem}{(\alph{mcitesubitemcount})}
\mciteSetBstSublistLabelBeginEnd
  {\mcitemaxwidthsubitemform\space}
  {\relax}
  {\relax}

\bibitem[Brandrup \latin{et~al.}(1999)Brandrup, Immergut, and
  Grulke]{Brandrup1999}
Brandrup,~J.; Immergut,~E.~H.; Grulke,~E.~A. \emph{Polymer Handbook};
  Wiley-Interscience: New York, 1999; p 2250\relax
\mciteBstWouldAddEndPuncttrue
\mciteSetBstMidEndSepPunct{\mcitedefaultmidpunct}
{\mcitedefaultendpunct}{\mcitedefaultseppunct}\relax
\EndOfBibitem
\end{mcitethebibliography}


\providecommand{\latin}[1]{#1}
\makeatletter
\providecommand{\doi}
  {\begingroup\let\do\@makeother\dospecials
  \catcode`\{=1 \catcode`\}=2 \doi@aux}
\providecommand{\doi@aux}[1]{\endgroup\texttt{#1}}
\makeatother
\providecommand*\mcitethebibliography{\thebibliography}
\csname @ifundefined\endcsname{endmcitethebibliography}
  {\let\endmcitethebibliography\endthebibliography}{}
\begin{mcitethebibliography}{61}
\providecommand*\natexlab[1]{#1}
\providecommand*\mciteSetBstSublistMode[1]{}
\providecommand*\mciteSetBstMaxWidthForm[2]{}
\providecommand*\mciteBstWouldAddEndPuncttrue
  {\def\EndOfBibitem{\unskip.}}
\providecommand*\mciteBstWouldAddEndPunctfalse
  {\let\EndOfBibitem\relax}
\providecommand*\mciteSetBstMidEndSepPunct[3]{}
\providecommand*\mciteSetBstSublistLabelBeginEnd[3]{}
\providecommand*\EndOfBibitem{}
\mciteSetBstSublistMode{f}
\mciteSetBstMaxWidthForm{subitem}{(\alph{mcitesubitemcount})}
\mciteSetBstSublistLabelBeginEnd
  {\mcitemaxwidthsubitemform\space}
  {\relax}
  {\relax}

\bibitem[Jana and Peng(2003)Jana, and Peng]{Jana2003}
Jana,~N.~R.; Peng,~X. Single-Phase and Gram-Scale Routes toward Nearly
  Monodisperse \ce{Au} and Other Noble Metal Nanocrystals. \emph{J. Am. Chem.
  Soc.} \textbf{2003}, \emph{125}, 14280--14281\relax
\mciteBstWouldAddEndPuncttrue
\mciteSetBstMidEndSepPunct{\mcitedefaultmidpunct}
{\mcitedefaultendpunct}{\mcitedefaultseppunct}\relax
\EndOfBibitem
\bibitem[Stoeva \latin{et~al.}(2002)Stoeva, Klabunde, Sorensen, and
  Dragieva]{Stoeva2002}
Stoeva,~S.; Klabunde,~K.~J.; Sorensen,~C.~M.; Dragieva,~I. Gram-Scale Synthesis
  of Monodisperse Gold Colloids by the Solvated Metal Atom Dispersion Method
  and Digestive Ripening and Their Organization into Two- and Three-Dimensional
  Structures. \emph{J. Am. Chem. Soc.} \textbf{2002}, \emph{124},
  2305--2311\relax
\mciteBstWouldAddEndPuncttrue
\mciteSetBstMidEndSepPunct{\mcitedefaultmidpunct}
{\mcitedefaultendpunct}{\mcitedefaultseppunct}\relax
\EndOfBibitem
\bibitem[Peng \latin{et~al.}(1998)Peng, Wickham, and Alivisatos]{Peng1998}
Peng,~X.; Wickham,~J.; Alivisatos,~A.~P. Kinetics of II-VI and III-V Colloidal
  Semiconductor Nanocrystal Growth: \enquote{Focusing} of Size Distributions.
  \emph{J. Am. Chem. Soc.} \textbf{1998}, \emph{120}, 5343--5344\relax
\mciteBstWouldAddEndPuncttrue
\mciteSetBstMidEndSepPunct{\mcitedefaultmidpunct}
{\mcitedefaultendpunct}{\mcitedefaultseppunct}\relax
\EndOfBibitem
\bibitem[Shim and Guyot-Sionnest(2000)Shim, and Guyot-Sionnest]{Shim2000}
Shim,~M.; Guyot-Sionnest,~P. \textit{n-}Type Colloidal Semiconductor
  Nanocrystals. \emph{Nature} \textbf{2000}, \emph{407}, 981--983\relax
\mciteBstWouldAddEndPuncttrue
\mciteSetBstMidEndSepPunct{\mcitedefaultmidpunct}
{\mcitedefaultendpunct}{\mcitedefaultseppunct}\relax
\EndOfBibitem
\bibitem[Pileni(2003)]{Pileni2003}
Pileni,~M.-P. The Role of Soft Colloidal Templates in Controlling the Size and
  Shape of Inorganic Nanocrystals. \emph{Nat. Mater.} \textbf{2003}, \emph{2},
  145--150\relax
\mciteBstWouldAddEndPuncttrue
\mciteSetBstMidEndSepPunct{\mcitedefaultmidpunct}
{\mcitedefaultendpunct}{\mcitedefaultseppunct}\relax
\EndOfBibitem
\bibitem[Murray \latin{et~al.}(1993)Murray, Norris, and Bawendi]{Murray1993}
Murray,~C.~B.; Norris,~D.~J.; Bawendi,~M.~G. Synthesis and Characterization of
  Nearly Monodisperse \ce{CdE} (\ce{E} = \ce{S}, \ce{Se}, \ce{Te})
  Semiconductor Nanocrystallites. \emph{J. Am. Chem. Soc.} \textbf{1993},
  \emph{115}, 8706--8715\relax
\mciteBstWouldAddEndPuncttrue
\mciteSetBstMidEndSepPunct{\mcitedefaultmidpunct}
{\mcitedefaultendpunct}{\mcitedefaultseppunct}\relax
\EndOfBibitem
\bibitem[Haruta(2003)]{Haruta2003}
Haruta,~M. When Gold Is Not Noble: Catalysis by Nanoparticles. \emph{Chem.
  Rec.} \textbf{2003}, \emph{3}, 75--87\relax
\mciteBstWouldAddEndPuncttrue
\mciteSetBstMidEndSepPunct{\mcitedefaultmidpunct}
{\mcitedefaultendpunct}{\mcitedefaultseppunct}\relax
\EndOfBibitem
\bibitem[Shipway \latin{et~al.}(2000)Shipway, Katz, and Willner]{Shipway2000}
Shipway,~A.~N.; Katz,~E.; Willner,~I. Nanoparticle Arrays on Surfaces for
  Electronic, Optical, and Sensor Applications. \emph{ChemPhysChem}
  \textbf{2000}, \emph{1}, 18--52\relax
\mciteBstWouldAddEndPuncttrue
\mciteSetBstMidEndSepPunct{\mcitedefaultmidpunct}
{\mcitedefaultendpunct}{\mcitedefaultseppunct}\relax
\EndOfBibitem
\bibitem[Saha \latin{et~al.}(2012)Saha, Agasti, Kim, Li, and Rotello]{Saha2012}
Saha,~K.; Agasti,~S.~S.; Kim,~C.; Li,~X.; Rotello,~V.~M. Gold Nanoparticles in
  Chemical and Biological Sensing. \emph{Chem. Rev.} \textbf{2012}, \emph{112},
  2739--2779\relax
\mciteBstWouldAddEndPuncttrue
\mciteSetBstMidEndSepPunct{\mcitedefaultmidpunct}
{\mcitedefaultendpunct}{\mcitedefaultseppunct}\relax
\EndOfBibitem
\bibitem[Talapin and Murray(2005)Talapin, and Murray]{Talapin2005}
Talapin,~D.~V.; Murray,~C.~B. PbSe Nanocrystal Solids for \textit{n-} and
  \textit{p-}Channel Thin Film Field-Effect Transistors. \emph{Science}
  \textbf{2005}, \emph{310}, 86--89\relax
\mciteBstWouldAddEndPuncttrue
\mciteSetBstMidEndSepPunct{\mcitedefaultmidpunct}
{\mcitedefaultendpunct}{\mcitedefaultseppunct}\relax
\EndOfBibitem
\bibitem[Zhao \latin{et~al.}(2015)Zhao, Rovere, Weerawarne, Osterhoudt, Kang,
  Joseph, Luo, Shim, Poliks, and Zhong]{Zhao2015}
Zhao,~W.; Rovere,~T.; Weerawarne,~D.; Osterhoudt,~G.; Kang,~N.; Joseph,~P.;
  Luo,~J.; Shim,~B.; Poliks,~M.; Zhong,~C.-J. Nanoalloy Printed and Pulse-Laser
  Sintered Flexible Sensor Devices with Enhanced Stability and Materials
  Compatibility. \emph{ACS Nano} \textbf{2015}, \emph{9}, 6168--6177\relax
\mciteBstWouldAddEndPuncttrue
\mciteSetBstMidEndSepPunct{\mcitedefaultmidpunct}
{\mcitedefaultendpunct}{\mcitedefaultseppunct}\relax
\EndOfBibitem
\bibitem[Chen \latin{et~al.}(2006)Chen, Hsu, and Hong]{Chen2006a}
Chen,~H.-S.; Hsu,~C.-K.; Hong,~H.-Y. \ce{InGaN}-\ce{CdSe}-\ce{ZnSe} Quantum
  Dots White LEDs. \emph{IEEE Photonics Technol. Lett.} \textbf{2006},
  \emph{18}, 193--195\relax
\mciteBstWouldAddEndPuncttrue
\mciteSetBstMidEndSepPunct{\mcitedefaultmidpunct}
{\mcitedefaultendpunct}{\mcitedefaultseppunct}\relax
\EndOfBibitem
\bibitem[Chen \latin{et~al.}(2006)Chen, Yeh, Lu, Huang, Shiao, Huang, Liu, and
  Su]{Chen2006b}
Chen,~H.-S.; Yeh,~D.-M.; Lu,~C.-F.; Huang,~C.-F.; Shiao,~W.-Y.;
  Huang,~C.~C.,~Jian-Jang amd~Yang; Liu,~I.-S.; Su,~W.-F. White Light
  Generation with \ce{CdSe}-\ce{ZnS} Nanocrystals Coated on an
  \ce{InGaN}-\ce{GaN} Quantum-Well Blue/Green Two-Wavelength Light-Emitting
  Diode. \emph{IEEE Photonics Technol. Lett.} \textbf{2006}, \emph{18},
  1430--1432\relax
\mciteBstWouldAddEndPuncttrue
\mciteSetBstMidEndSepPunct{\mcitedefaultmidpunct}
{\mcitedefaultendpunct}{\mcitedefaultseppunct}\relax
\EndOfBibitem
\bibitem[Achermann \latin{et~al.}(2006)Achermann, Petruska, Koleske, Crawford,
  and Klimov]{Achermann2006}
Achermann,~M.; Petruska,~M.~A.; Koleske,~D.~D.; Crawford,~M.~H.; Klimov,~V.~I.
  Nanocrystal-Based Light-Emitting Diodes Utilizing High-Efficiency
  Nonradiative Energy Transfer for Color Conversion. \emph{Nano Lett.}
  \textbf{2006}, \emph{6}, 1396--1400\relax
\mciteBstWouldAddEndPuncttrue
\mciteSetBstMidEndSepPunct{\mcitedefaultmidpunct}
{\mcitedefaultendpunct}{\mcitedefaultseppunct}\relax
\EndOfBibitem
\bibitem[Korgel and Fitzmaurice(1998)Korgel, and Fitzmaurice]{Korgel1998}
Korgel,~B.~A.; Fitzmaurice,~D. Condensation of Ordered Nanocrystal Thin Films.
  \emph{Phys. Rev. Lett.} \textbf{1998}, \emph{80}, 3531--3534\relax
\mciteBstWouldAddEndPuncttrue
\mciteSetBstMidEndSepPunct{\mcitedefaultmidpunct}
{\mcitedefaultendpunct}{\mcitedefaultseppunct}\relax
\EndOfBibitem
\bibitem[Brust \latin{et~al.}(1994)Brust, Walker, Bethell, Schiffrin, and
  Whyman]{Brust1994}
Brust,~M.; Walker,~M.; Bethell,~D.; Schiffrin,~D.~J.; Whyman,~R. Synthesis of
  Thiol-Derivatised Gold Nanoparticles in a Two-Phase Liquid{\textendash}Liquid
  System. \emph{J. Chem. Soc., Chem. Commun.} \textbf{1994}, \emph{0},
  801--802\relax
\mciteBstWouldAddEndPuncttrue
\mciteSetBstMidEndSepPunct{\mcitedefaultmidpunct}
{\mcitedefaultendpunct}{\mcitedefaultseppunct}\relax
\EndOfBibitem
\bibitem[Puntes \latin{et~al.}(2001)Puntes, Krishnan, and
  Alivisatos]{Puntes2001}
Puntes,~V.~F.; Krishnan,~K.~M.; Alivisatos,~A.~P. Colloidal Nanocrystal Shape
  and Size Control: The Case of Cobalt. \emph{Science} \textbf{2001},
  \emph{291}, 2115--2117\relax
\mciteBstWouldAddEndPuncttrue
\mciteSetBstMidEndSepPunct{\mcitedefaultmidpunct}
{\mcitedefaultendpunct}{\mcitedefaultseppunct}\relax
\EndOfBibitem
\bibitem[Napper(1977)]{Napper1977}
Napper,~D. Steric Stabilization. \emph{J. Colloid Interface Sci.}
  \textbf{1977}, \emph{58}, 390--407\relax
\mciteBstWouldAddEndPuncttrue
\mciteSetBstMidEndSepPunct{\mcitedefaultmidpunct}
{\mcitedefaultendpunct}{\mcitedefaultseppunct}\relax
\EndOfBibitem
\bibitem[Sigman \latin{et~al.}(2004)Sigman, Saunders, and Korgel]{Sigman2004}
Sigman,~M.~B.; Saunders,~A.~E.; Korgel,~B.~A. Metal Nanocrystal Superlattice
  Nucleation and Growth. \emph{Langmuir} \textbf{2004}, \emph{20},
  978--983\relax
\mciteBstWouldAddEndPuncttrue
\mciteSetBstMidEndSepPunct{\mcitedefaultmidpunct}
{\mcitedefaultendpunct}{\mcitedefaultseppunct}\relax
\EndOfBibitem
\bibitem[Bodnarchuk \latin{et~al.}(2010)Bodnarchuk, Kovalenko, Heiss, and
  Talapin]{Bodnarchuk2010}
Bodnarchuk,~M.~I.; Kovalenko,~M.~V.; Heiss,~W.; Talapin,~D.~V. Energetic and
  Entropic Contributions to Self-Assembly of Binary Nanocrystal Superlattices:
  Temperature as the Structure-Directing Factor. \emph{J. Am. Chem. Soc.}
  \textbf{2010}, \emph{132}, 11967--11977\relax
\mciteBstWouldAddEndPuncttrue
\mciteSetBstMidEndSepPunct{\mcitedefaultmidpunct}
{\mcitedefaultendpunct}{\mcitedefaultseppunct}\relax
\EndOfBibitem
\bibitem[Schapotschnikow \latin{et~al.}(2008)Schapotschnikow, Pool, and
  Vlugt]{Schapotschnikow2008}
Schapotschnikow,~P.; Pool,~R.; Vlugt,~T. J.~H. Molecular Simulations of
  Interacting Nanocrystals. \emph{Nano Lett.} \textbf{2008}, \emph{8},
  2930--2934\relax
\mciteBstWouldAddEndPuncttrue
\mciteSetBstMidEndSepPunct{\mcitedefaultmidpunct}
{\mcitedefaultendpunct}{\mcitedefaultseppunct}\relax
\EndOfBibitem
\bibitem[Patel and Egorov(2007)Patel, and Egorov]{Patel2007}
Patel,~N.; Egorov,~S.~A. Interactions between Sterically Stabilized
  Nanoparticles in Supercritical Fluids: A Simulation Study. \emph{J. Chem.
  Phys.} \textbf{2007}, \emph{126}, 054706\relax
\mciteBstWouldAddEndPuncttrue
\mciteSetBstMidEndSepPunct{\mcitedefaultmidpunct}
{\mcitedefaultendpunct}{\mcitedefaultseppunct}\relax
\EndOfBibitem
\bibitem[Khan \latin{et~al.}(2009)Khan, Pierce, Sorensen, and
  Chakrabarti]{Khan2009}
Khan,~S.~J.; Pierce,~F.; Sorensen,~C.~M.; Chakrabarti,~A. Self-Assembly of
  Ligated Gold Nanoparticles: Phenomenological Modeling and Computer
  Simulations. \emph{Langmuir} \textbf{2009}, \emph{25}, 13861--13868\relax
\mciteBstWouldAddEndPuncttrue
\mciteSetBstMidEndSepPunct{\mcitedefaultmidpunct}
{\mcitedefaultendpunct}{\mcitedefaultseppunct}\relax
\EndOfBibitem
\bibitem[Dalmaschio \latin{et~al.}(2013)Dalmaschio, da~Silveira~Firmiano,
  Pinheiro, Sobrinho, Farias~de Moura, and Leite]{Dalmaschio2013}
Dalmaschio,~C.~J.; da~Silveira~Firmiano,~E.~G.; Pinheiro,~A.~N.;
  Sobrinho,~D.~G.; Farias~de Moura,~A.; Leite,~E.~R. Nanocrystals
  Self-Assembled in Superlattices Directed by the Solvent-Organic Capping
  Interaction. \emph{Nanoscale} \textbf{2013}, \emph{5}, 5602--10\relax
\mciteBstWouldAddEndPuncttrue
\mciteSetBstMidEndSepPunct{\mcitedefaultmidpunct}
{\mcitedefaultendpunct}{\mcitedefaultseppunct}\relax
\EndOfBibitem
\bibitem[Widmer-Cooper and Geissler(2014)Widmer-Cooper, and
  Geissler]{Widmer-Cooper2014}
Widmer-Cooper,~A.; Geissler,~P. Orientational Ordering of Passivating Ligands
  on \ce{CdS} Nanorods in Solution Generates Strong Rod-Rod Interactions.
  \emph{Nano Lett.} \textbf{2014}, \emph{14}, 57--65\relax
\mciteBstWouldAddEndPuncttrue
\mciteSetBstMidEndSepPunct{\mcitedefaultmidpunct}
{\mcitedefaultendpunct}{\mcitedefaultseppunct}\relax
\EndOfBibitem
\bibitem[Moura \latin{et~al.}(2015)Moura, Bernardino, Dalmaschio, Leite, and
  Kotov]{Moura2015}
Moura,~F.~D.; Bernardino,~K.; Dalmaschio,~C.~J.; Leite,~E.~R.; Kotov,~N.~A.
  {Thermodynamic Insights into the Self-Assembly of Capped Nanoparticles Using
  Molecular Dynamic}. \emph{Phys. Chem. Chem. Phys.} \textbf{2015}, \emph{17},
  3820--3831\relax
\mciteBstWouldAddEndPuncttrue
\mciteSetBstMidEndSepPunct{\mcitedefaultmidpunct}
{\mcitedefaultendpunct}{\mcitedefaultseppunct}\relax
\EndOfBibitem
\bibitem[Talapin \latin{et~al.}(2001)Talapin, Shevchenko, Kornowski, Gaponik,
  Haase, Rogach, and Weller]{Talapin2001}
Talapin,~D.~V.; Shevchenko,~E.~V.; Kornowski,~A.; Gaponik,~N.; Haase,~M.;
  Rogach,~A.~L.; Weller,~H. A New Approach to Crystallization of \ce{CdSe}
  Nanoparticles into Ordered Three-Dimensional Superlattices. \emph{Adv.
  Mater.} \textbf{2001}, \emph{13}, 1868\relax
\mciteBstWouldAddEndPuncttrue
\mciteSetBstMidEndSepPunct{\mcitedefaultmidpunct}
{\mcitedefaultendpunct}{\mcitedefaultseppunct}\relax
\EndOfBibitem
\bibitem[Ab\'ecassis \latin{et~al.}(2008)Ab\'ecassis, Testard, and
  Spalla]{Abecassis2008}
Ab\'ecassis,~B.; Testard,~F.; Spalla,~O. Gold Nanoparticle Superlattice
  Crystallization Probed \textit{In Situ}. \emph{Phys. Rev. Lett.}
  \textbf{2008}, \emph{100}, 115504\relax
\mciteBstWouldAddEndPuncttrue
\mciteSetBstMidEndSepPunct{\mcitedefaultmidpunct}
{\mcitedefaultendpunct}{\mcitedefaultseppunct}\relax
\EndOfBibitem
\bibitem[Khan \latin{et~al.}(2012)Khan, Sorensen, and Chakrabarti]{Khan2012}
Khan,~S.~J.; Sorensen,~C.~M.; Chakrabarti,~A. Computer Simulations of
  Nucleation of Nanoparticle Superclusters from Solution. \emph{Langmuir}
  \textbf{2012}, \emph{28}, 5570--5579\relax
\mciteBstWouldAddEndPuncttrue
\mciteSetBstMidEndSepPunct{\mcitedefaultmidpunct}
{\mcitedefaultendpunct}{\mcitedefaultseppunct}\relax
\EndOfBibitem
\bibitem[Kister \latin{et~al.}(2018)Kister, Monego, Mulvaney, Widmer-Cooper,
  and Kraus]{Kister2018}
Kister,~T.; Monego,~D.; Mulvaney,~P.; Widmer-Cooper,~A.; Kraus,~T. Colloidal
  Stability of Apolar Nanoparticles: The Role of Particle Size and Ligand Shell
  Structure. \emph{ACS Nano} \textbf{2018}, \emph{12}, 5969--5977\relax
\mciteBstWouldAddEndPuncttrue
\mciteSetBstMidEndSepPunct{\mcitedefaultmidpunct}
{\mcitedefaultendpunct}{\mcitedefaultseppunct}\relax
\EndOfBibitem
\bibitem[Monego \latin{et~al.}(2018)Monego, Kister, Kirkwood, Mulvaney,
  Widmer-Cooper, and Kraus]{Monego2018}
Monego,~D.; Kister,~T.; Kirkwood,~N.; Mulvaney,~P.; Widmer-Cooper,~A.;
  Kraus,~T. Colloidal Stability of Apolar Nanoparticles: Role of Ligand Length.
  \emph{Langmuir} \textbf{2018}, \emph{34}, 12982--12989\relax
\mciteBstWouldAddEndPuncttrue
\mciteSetBstMidEndSepPunct{\mcitedefaultmidpunct}
{\mcitedefaultendpunct}{\mcitedefaultseppunct}\relax
\EndOfBibitem
\bibitem[Lohman \latin{et~al.}(2012)Lohman, Powell, Cingarapu, Aakeroy,
  Chakrabarti, Klabunde, Law, and Sorensen]{Lohman2012}
Lohman,~B.~C.; Powell,~J.~A.; Cingarapu,~S.; Aakeroy,~C.~B.; Chakrabarti,~A.;
  Klabunde,~K.~J.; Law,~B.~M.; Sorensen,~C.~M. Solubility of Gold Nanoparticles
  as a Function of Ligand Shell and Alkane Solvent. \emph{Phys. Chem. Chem.
  Phys.} \textbf{2012}, \emph{14}, 6509--6513\relax
\mciteBstWouldAddEndPuncttrue
\mciteSetBstMidEndSepPunct{\mcitedefaultmidpunct}
{\mcitedefaultendpunct}{\mcitedefaultseppunct}\relax
\EndOfBibitem
\bibitem[Hajiw \latin{et~al.}(2015)Hajiw, Schmitt, Imperor-Clerc, and
  Pansu]{Hajiw2015}
Hajiw,~S.; Schmitt,~J.; Imperor-Clerc,~M.; Pansu,~B. {Solvent-Driven
  Interactions between Hydrophobically-Coated Nanoparticles}. \emph{Soft
  Matter} \textbf{2015}, \emph{11}, 3920--3926\relax
\mciteBstWouldAddEndPuncttrue
\mciteSetBstMidEndSepPunct{\mcitedefaultmidpunct}
{\mcitedefaultendpunct}{\mcitedefaultseppunct}\relax
\EndOfBibitem
\bibitem[de~Gennes(1980)]{deGennes1980}
de~Gennes,~P.~G. Conformations of Polymers Attached to an Interface.
  \emph{Macromolecules} \textbf{1980}, \emph{13}, 1069--1075\relax
\mciteBstWouldAddEndPuncttrue
\mciteSetBstMidEndSepPunct{\mcitedefaultmidpunct}
{\mcitedefaultendpunct}{\mcitedefaultseppunct}\relax
\EndOfBibitem
\bibitem[Koski \latin{et~al.}(2019)Koski, Krook, Ford, Yahata, Ohno, Murray,
  Frischknecht, Composto, and Riggleman]{Koski2019}
Koski,~J.~P.; Krook,~N.~M.; Ford,~J.; Yahata,~Y.; Ohno,~K.; Murray,~C.~B.;
  Frischknecht,~A.~L.; Composto,~R.~J.; Riggleman,~R.~A. Phase Behavior of
  Grafted Polymer Nanocomposites from Field-Based Simulations.
  \emph{Macromolecules} \textbf{2019}, \emph{52}, 5110--5121\relax
\mciteBstWouldAddEndPuncttrue
\mciteSetBstMidEndSepPunct{\mcitedefaultmidpunct}
{\mcitedefaultendpunct}{\mcitedefaultseppunct}\relax
\EndOfBibitem
\bibitem[Sunday \latin{et~al.}(2012)Sunday, Ilavsky, and Green]{Sunday2012}
Sunday,~D.; Ilavsky,~J.; Green,~D.~L. A Phase Diagram for Polymer-Grafted
  Nanoparticles in Homopolymer Matrices. \emph{Macromolecules} \textbf{2012},
  \emph{45}, 4007--4011\relax
\mciteBstWouldAddEndPuncttrue
\mciteSetBstMidEndSepPunct{\mcitedefaultmidpunct}
{\mcitedefaultendpunct}{\mcitedefaultseppunct}\relax
\EndOfBibitem
\bibitem[Sunday and Green(2015)Sunday, and Green]{Sunday2015}
Sunday,~D.~F.; Green,~D.~L. Thermal and Rheological Behavior of Polymer Grafted
  Nanoparticles. \emph{Macromolecules} \textbf{2015}, \emph{48},
  8651--8659\relax
\mciteBstWouldAddEndPuncttrue
\mciteSetBstMidEndSepPunct{\mcitedefaultmidpunct}
{\mcitedefaultendpunct}{\mcitedefaultseppunct}\relax
\EndOfBibitem
\bibitem[Zou \latin{et~al.}(2017)Zou, Ban, Cui, Huang, Wu, Liu, and
  Wu]{Zou2017}
Zou,~Y.; Ban,~M.; Cui,~W.; Huang,~Q.; Wu,~C.; Liu,~J.; Wu,~H. {A General
  Solvent Selection Strategy for Solution Processed Quantum Dots Targeting High
  Performance Light-Emitting Diode}. \emph{Adv. Funct. Mater.} \textbf{2017},
  \emph{27}, 1603325\relax
\mciteBstWouldAddEndPuncttrue
\mciteSetBstMidEndSepPunct{\mcitedefaultmidpunct}
{\mcitedefaultendpunct}{\mcitedefaultseppunct}\relax
\EndOfBibitem
\bibitem[Fini \latin{et~al.}(2008)Fini, Depalo, Comparelli, Curri, Striccoli,
  Castagnolo, and Agostiano]{Fini2008}
Fini,~P.; Depalo,~N.; Comparelli,~R.; Curri,~M.~L.; Striccoli,~M.;
  Castagnolo,~M.; Agostiano,~A. Interactions between Surfactant Capped \ce{CdS}
  Nanocrystals and Organic Solvent. \emph{J. Therm. Anal. Calorim.}
  \textbf{2008}, \emph{92}, 271--277\relax
\mciteBstWouldAddEndPuncttrue
\mciteSetBstMidEndSepPunct{\mcitedefaultmidpunct}
{\mcitedefaultendpunct}{\mcitedefaultseppunct}\relax
\EndOfBibitem
\bibitem[Widmer-Cooper and Geissler(2016)Widmer-Cooper, and
  Geissler]{Widmer-Cooper2016}
Widmer-Cooper,~A.; Geissler,~P.~L. Ligand-Mediated Interactions between
  Nanoscale Surfaces Depend Sensitively and Nonlinearly on Temperature, Facet
  Dimensions, and Ligand Coverage. \emph{ACS Nano} \textbf{2016}, \emph{10},
  1877--87\relax
\mciteBstWouldAddEndPuncttrue
\mciteSetBstMidEndSepPunct{\mcitedefaultmidpunct}
{\mcitedefaultendpunct}{\mcitedefaultseppunct}\relax
\EndOfBibitem
\bibitem[Johnson(1959)]{Johnson1959}
Johnson,~J.~E. X-Ray Diffraction Studies of the Crystallinity in Polyethylene
  Terephthalate. \emph{J. Appl. Polym. Sci.} \textbf{1959}, \emph{2},
  205--209\relax
\mciteBstWouldAddEndPuncttrue
\mciteSetBstMidEndSepPunct{\mcitedefaultmidpunct}
{\mcitedefaultendpunct}{\mcitedefaultseppunct}\relax
\EndOfBibitem
\bibitem[Brandrup \latin{et~al.}(1999)Brandrup, Immergut, and
  Grulke]{Brandrup1999}
Brandrup,~J.; Immergut,~E.~H.; Grulke,~E.~A. \emph{Polymer Handbook};
  Wiley-Interscience: New York, 1999; p 2250\relax
\mciteBstWouldAddEndPuncttrue
\mciteSetBstMidEndSepPunct{\mcitedefaultmidpunct}
{\mcitedefaultendpunct}{\mcitedefaultseppunct}\relax
\EndOfBibitem
\bibitem[Costa \latin{et~al.}(2018)Costa, Mendes, and Santos]{Costa2018}
Costa,~J.; Mendes,~A.; Santos,~L. {Chain Length Dependence of the Thermodynamic
  Properties of \textit{n-}Alkanes and their Monosubstituted Derivatives}.
  \emph{J. Chem. Eng. Data} \textbf{2018}, \emph{63}, 1\relax
\mciteBstWouldAddEndPuncttrue
\mciteSetBstMidEndSepPunct{\mcitedefaultmidpunct}
{\mcitedefaultendpunct}{\mcitedefaultseppunct}\relax
\EndOfBibitem
\bibitem[Roke \latin{et~al.}(2006)Roke, Berg, Buitenhuis, Blaaderen, and
  Bonn]{Roke2006}
Roke,~S.; Berg,~O.; Buitenhuis,~J.; Blaaderen,~A.~V.; Bonn,~M. {Surface
  Molecular View of Colloidal Gelation}. \emph{Proc. Natl. Acad. Sci. USA}
  \textbf{2006}, \emph{103}, 13310--13314\relax
\mciteBstWouldAddEndPuncttrue
\mciteSetBstMidEndSepPunct{\mcitedefaultmidpunct}
{\mcitedefaultendpunct}{\mcitedefaultseppunct}\relax
\EndOfBibitem
\bibitem[Roo \latin{et~al.}(2018)Roo, Yazdani, Drijvers, Lauria, Maes, Owen,
  Driessche, Niederberger, Wood, Martins, Infante, and Hens]{Roo2018}
Roo,~J.~D.; Yazdani,~N.; Drijvers,~E.; Lauria,~A.; Maes,~J.; Owen,~J.~S.;
  Driessche,~I.~V.; Niederberger,~M.; Wood,~V.; Martins,~J.~C.; Infante,~I.;
  Hens,~Z. {Probing Solvent-Ligand Interactions in Colloidal Nanocrystals by
  the NMR Line Broadening}. \emph{Chem. Mater.} \textbf{2018}, \emph{30},
  5485--5492\relax
\mciteBstWouldAddEndPuncttrue
\mciteSetBstMidEndSepPunct{\mcitedefaultmidpunct}
{\mcitedefaultendpunct}{\mcitedefaultseppunct}\relax
\EndOfBibitem
\bibitem[Ruks \latin{et~al.}(2019)Ruks, Beuck, Schaller, Niemeyer,
  Z{\"{a}}hres, Loza, Heggen, Hagemann, Mayer, Bayer, and Epple]{Ruks2019}
Ruks,~T.; Beuck,~C.; Schaller,~T.; Niemeyer,~F.; Z{\"{a}}hres,~M.; Loza,~K.;
  Heggen,~M.; Hagemann,~U.; Mayer,~C.; Bayer,~P.; Epple,~M. {Solution NMR
  Spectroscopy with Isotope-Labeled Cysteine (13C and 15N) Reveals the Surface
  Structure of l-Cysteine-Coated Ultrasmall Gold Nanoparticles
  (\SI{1.8}{\nano\meter})}. \emph{Langmuir} \textbf{2019}, \emph{35},
  767--778\relax
\mciteBstWouldAddEndPuncttrue
\mciteSetBstMidEndSepPunct{\mcitedefaultmidpunct}
{\mcitedefaultendpunct}{\mcitedefaultseppunct}\relax
\EndOfBibitem
\bibitem[Luo \latin{et~al.}(2018)Luo, Marson, Ong, Loiudice, Kohlbrecher,
  Radulescu, Krause-Heuer, Darwish, Balog, Buonsanti, Svergun, Posocco, and
  Stellacci]{Luo2018}
Luo,~Z.; Marson,~D.; Ong,~Q.~K.; Loiudice,~A.; Kohlbrecher,~J.; Radulescu,~A.;
  Krause-Heuer,~A.; Darwish,~T.; Balog,~S.; Buonsanti,~R.; Svergun,~D.~I.;
  Posocco,~P.; Stellacci,~F. {Quantitative 3D Determination of Self-Assembled
  Structures on Nanoparticles Using Small Angle Neutron Scattering}. \emph{Nat.
  Commun.} \textbf{2018}, \emph{9}, 1343\relax
\mciteBstWouldAddEndPuncttrue
\mciteSetBstMidEndSepPunct{\mcitedefaultmidpunct}
{\mcitedefaultendpunct}{\mcitedefaultseppunct}\relax
\EndOfBibitem
\bibitem[Lane and Grest(2010)Lane, and Grest]{Lane2010}
Lane,~J. M.~D.; Grest,~G.~S. {Spontaneous Asymmetry of Coated Spherical
  Nanoparticles in Solution and at Liquid-Vapor Interfaces}. \emph{Phys. Rev.
  Lett.} \textbf{2010}, \emph{104}, 235501\relax
\mciteBstWouldAddEndPuncttrue
\mciteSetBstMidEndSepPunct{\mcitedefaultmidpunct}
{\mcitedefaultendpunct}{\mcitedefaultseppunct}\relax
\EndOfBibitem
\bibitem[Bolintineanu \latin{et~al.}(2014)Bolintineanu, Lane, and
  Grest]{Bolintineanu2014}
Bolintineanu,~D.~S.; Lane,~J. M.~D.; Grest,~G.~S. {Effects of Functional Groups
  and Ionization on the Structure of Alkanethiol-Coated Gold Nanoparticles}.
  \emph{Langmuir} \textbf{2014}, \emph{30}, 11075--11085\relax
\mciteBstWouldAddEndPuncttrue
\mciteSetBstMidEndSepPunct{\mcitedefaultmidpunct}
{\mcitedefaultendpunct}{\mcitedefaultseppunct}\relax
\EndOfBibitem
\bibitem[Kister \latin{et~al.}(2016)Kister, Mravlak, Schilling, and
  Kraus]{Kister2016pressure}
Kister,~T.; Mravlak,~M.; Schilling,~T.; Kraus,~T. Pressure-Controlled Formation
  of Crystalline, Janus, and Core-Shell Supraparticles. \emph{Nanoscale}
  \textbf{2016}, \emph{8}, 13377--13384\relax
\mciteBstWouldAddEndPuncttrue
\mciteSetBstMidEndSepPunct{\mcitedefaultmidpunct}
{\mcitedefaultendpunct}{\mcitedefaultseppunct}\relax
\EndOfBibitem
\bibitem[Chambrier \latin{et~al.}(2015)Chambrier, Banerjee,
  Remiro-Buenamañana, Chao, Cammidge, and Bochmann]{Chambrier2015}
Chambrier,~I.; Banerjee,~C.; Remiro-Buenamañana,~S.; Chao,~Y.;
  Cammidge,~A.~N.; Bochmann,~M. Synthesis of Porphyrin-\ce{CdSe} Quantum Dot
  Assemblies: Controlling Ligand Binding by Substituent Effects. \emph{Inorg.
  Chem.} \textbf{2015}, \emph{54}, 7368--7380\relax
\mciteBstWouldAddEndPuncttrue
\mciteSetBstMidEndSepPunct{\mcitedefaultmidpunct}
{\mcitedefaultendpunct}{\mcitedefaultseppunct}\relax
\EndOfBibitem
\bibitem[Schnablegger and Singh(2013)Schnablegger, and
  Singh]{Schnablegger2013saxs}
Schnablegger,~H.; Singh,~Y. \emph{The SAXS Guide: Getting Acquainted with the
  Principles}; Anton Paar GmbH: Graz, Austria, 2013; Vol.~2\relax
\mciteBstWouldAddEndPuncttrue
\mciteSetBstMidEndSepPunct{\mcitedefaultmidpunct}
{\mcitedefaultendpunct}{\mcitedefaultseppunct}\relax
\EndOfBibitem
\bibitem[Andersen(1983)]{Andersen1983}
Andersen,~H.~C. Rattle: A “Velocity” Version of the Shake Algorithm for
  Molecular Dynamics Calculations. \emph{J. Comput. Phys.} \textbf{1983},
  \emph{52}, 24--34\relax
\mciteBstWouldAddEndPuncttrue
\mciteSetBstMidEndSepPunct{\mcitedefaultmidpunct}
{\mcitedefaultendpunct}{\mcitedefaultseppunct}\relax
\EndOfBibitem
\bibitem[Pool \latin{et~al.}(2007)Pool, Schapotschnikow, and Vlugt]{Pool2007}
Pool,~R.; Schapotschnikow,~P.; Vlugt,~T. J.~H. Solvent Effects in the
  Adsorption of Alkyl Thiols on Gold Structures: A Molecular Simulation Study.
  \emph{J. Phys. Chem. C} \textbf{2007}, \emph{111}, 10201--10212\relax
\mciteBstWouldAddEndPuncttrue
\mciteSetBstMidEndSepPunct{\mcitedefaultmidpunct}
{\mcitedefaultendpunct}{\mcitedefaultseppunct}\relax
\EndOfBibitem
\bibitem[Martin and Siepmann(1998)Martin, and Siepmann]{Martin1998}
Martin,~M.~G.; Siepmann,~J.~I. Transferable Potentials for Phase Equilibria. 1.
  United-Atom Description of \textit{n-}Alkanes. \emph{J. Phys. Chem. B}
  \textbf{1998}, \emph{102}, 2569--2577\relax
\mciteBstWouldAddEndPuncttrue
\mciteSetBstMidEndSepPunct{\mcitedefaultmidpunct}
{\mcitedefaultendpunct}{\mcitedefaultseppunct}\relax
\EndOfBibitem
\bibitem[Plimpton(1995)]{Plimpton1995}
Plimpton,~S. Fast Parallel Algorithms for Short-Range Molecular Dynamics.
  \emph{J. Comput. Phys.} \textbf{1995}, \emph{117}, 1--19\relax
\mciteBstWouldAddEndPuncttrue
\mciteSetBstMidEndSepPunct{\mcitedefaultmidpunct}
{\mcitedefaultendpunct}{\mcitedefaultseppunct}\relax
\EndOfBibitem
\bibitem[Humphrey \latin{et~al.}(1996)Humphrey, Dalke, and
  Schulten]{Humphrey1996}
Humphrey,~W.; Dalke,~A.; Schulten,~K. VMD: Visual Molecular Dynamics. \emph{J.
  Mol. Graphics} \textbf{1996}, \emph{14}, 33--38\relax
\mciteBstWouldAddEndPuncttrue
\mciteSetBstMidEndSepPunct{\mcitedefaultmidpunct}
{\mcitedefaultendpunct}{\mcitedefaultseppunct}\relax
\EndOfBibitem
\bibitem[Hamaker(1937)]{Hamaker1937}
Hamaker,~H.~C. The London-van der Waals Attraction between Spherical Particles.
  \emph{Physica} \textbf{1937}, \emph{4}, 1058--1072\relax
\mciteBstWouldAddEndPuncttrue
\mciteSetBstMidEndSepPunct{\mcitedefaultmidpunct}
{\mcitedefaultendpunct}{\mcitedefaultseppunct}\relax
\EndOfBibitem
\bibitem[Ederth(2001)]{Ederth2001}
Ederth,~T. Computation of Lifshitz-van der Waals Forces between Alkylthiol
  Monolayers on Gold Films. \emph{Langmuir} \textbf{2001}, \emph{17},
  3329--3340\relax
\mciteBstWouldAddEndPuncttrue
\mciteSetBstMidEndSepPunct{\mcitedefaultmidpunct}
{\mcitedefaultendpunct}{\mcitedefaultseppunct}\relax
\EndOfBibitem
\bibitem[Su{\'{a}}rez \latin{et~al.}(2011)Su{\'{a}}rez, D{\'{\i}}az, and
  Su{\'{a}}rez]{Suarez2011}
Su{\'{a}}rez,~E.; D{\'{\i}}az,~N.; Su{\'{a}}rez,~D. Entropy Calculations of
  Single Molecules by Combining the Rigid{\textendash}Rotor and
  Harmonic-Oscillator Approximations with Conformational Entropy Estimations
  from Molecular Dynamics Simulations. \emph{J. Chem. Theory Comput.}
  \textbf{2011}, \emph{7}, 2638--2653\relax
\mciteBstWouldAddEndPuncttrue
\mciteSetBstMidEndSepPunct{\mcitedefaultmidpunct}
{\mcitedefaultendpunct}{\mcitedefaultseppunct}\relax
\EndOfBibitem
\end{mcitethebibliography}
\end{document}

\begin{figure}[H]
    \centering
    \includegraphics[width=\linewidth]{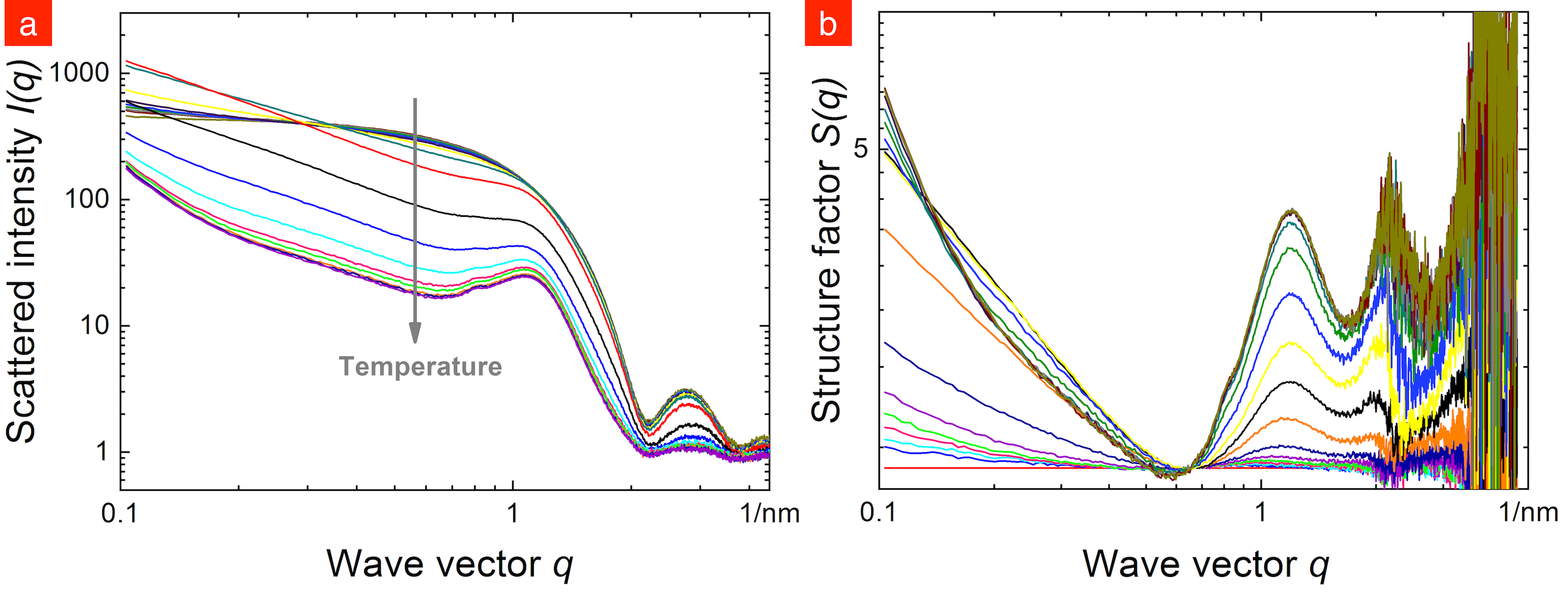}
    \caption{Temperature-dependent scattering of \SI{4}{\nano\meter} \ce{Au} nanoparticles. a) Raw data. b) The calculated structure factors. The peaks increase upon cooling.} 
    \label{SI_SAXS_data}
    \vspace{0.5em}
\end{figure}

Figure \ref{SI_SAXS_data} shows the scattering curves of AuNPs and the structure factor contributions during a cooling cycle. The development of the first peak in the structure factor indicates the agglomeration of the nanoparticles.

\begin{figure}[H]
    \centering
    \includegraphics[width=0.8\linewidth]{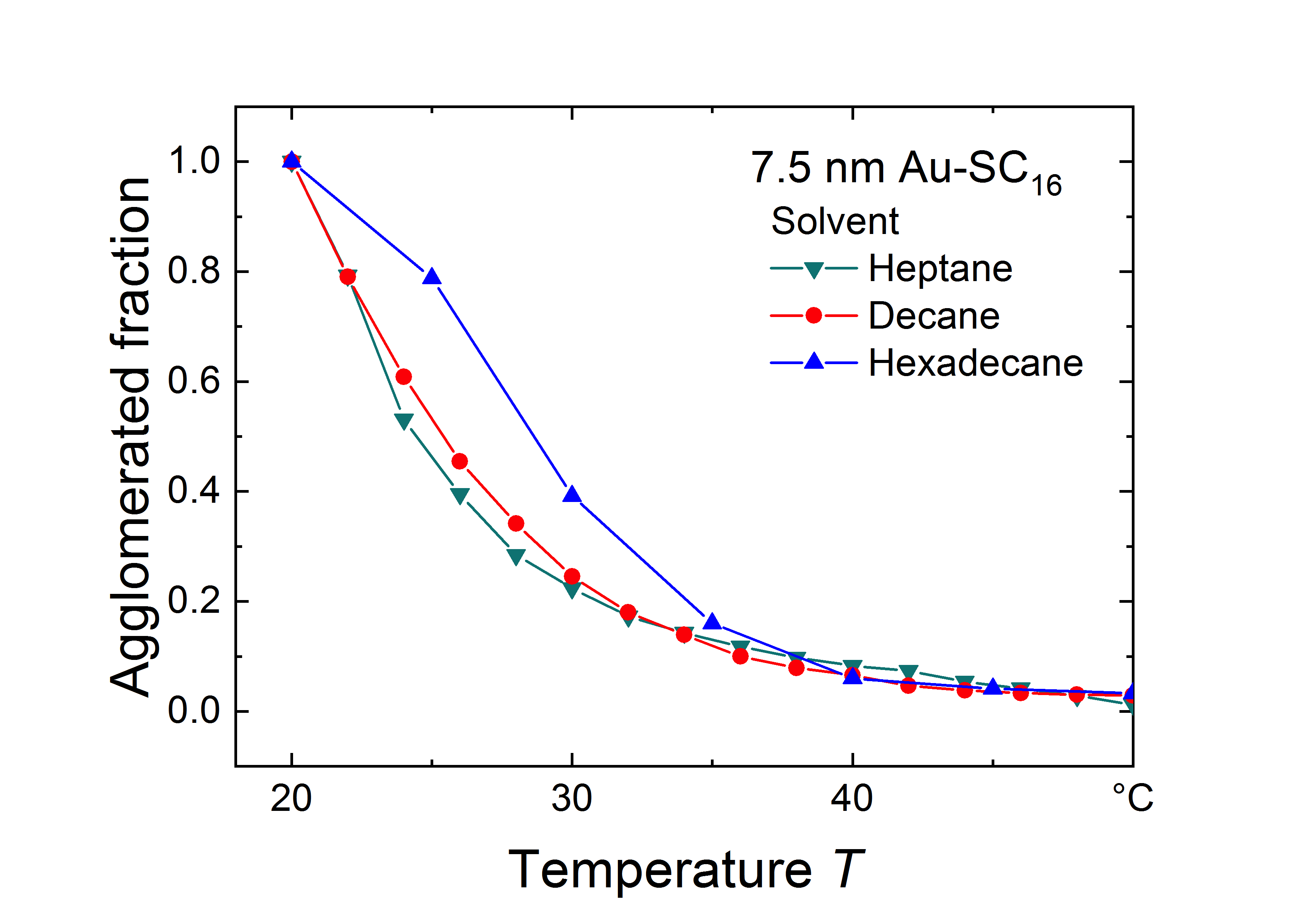}
    \caption{Fraction of agglomerated \SI{7.5}{\nano\meter} \ce{Au-SC_{16}} particles as determined by \emph{in situ} small-angle X-ray scattering.} 
    \label{SI_7p5nmSC16}
    \vspace{0.5em}
\end{figure}

 The trend observed for the agglomeration temperature when changing the length of the alkane in which they are dispersed is the same for particles with a larger core diameter: \ce{Au} particles of \SI{7.5}{\nano\meter} in diameter coated with hexadecanethiol ligands (Figure \ref{SI_7p5nmSC16}).

Table \ref{SI_Table_SolvParams} shows the solubility parameters, from both Hildebrand and Hansen theories, for the solvents studied here.

\begin{table}[H]

\begin{adjustbox}{width=\textwidth}

\begin{tabular}{@{}|cc|ccccc|c|@{}}
\hline
\multicolumn{2}{|c|}{Ligand} & \multicolumn{5}{c|}{Solvent} & \multirow{3}{*}{Flory Parameter} \\ \cline{1-7} & \multirow{2}{*}{Hildebrand Parameter ($Pa^{1/2}$)\cite{Brandrup1999}} &  & \multirow{2}{*}{Hildebrand Parameter ($Pa^{1/2}$)\cite{Brandrup1999}} & \multicolumn{3}{c|}{Hansen Parameter\cite{Brandrup1999}} &  \\  & & & & dispersion & dipolar & $H_{bond}$ & \\ \hline 
\multirow{4}{*}{\ce{SC_{16}}} & \multirow{4}{*}{16400} & Hexane & 14100 & 14100 & 0 & 0 & 0.62 \\
& & Decane & 15800 & 15800 & 0 & 0 & 0.37 \\
& & Hexadecane & 16400 & 16400 & 0 & 0 & 0.34 \\ 
& & Cyclohexane & 16800 & 16800 & 0 & 200 & 0.35 \\ \hline
\multirow{4}{*}{\ce{SC_{18}}} & \multirow{4}{*}{17100} & Hexane & 14100 & 14100 & 0 & 0 & 0.82 \\
& & Decane & 15800 & 15800 & 0 & 0 & 0.47 \\
& & Hexadecane & 16400 & 16400 & 0 & 0 & 0.40 \\ 
& & Cyclohexane & 16800 & 16800 & 0 & 200 & 0.34 \\ \hline 
\end{tabular}

\end{adjustbox}
\caption{Hildebrand and Hansen solubility parameters for the solvents studied here and the ligand-solvent interaction Flory parameter ($\chi$), calculated as per Equation 1. Solubility parameters for the ligands were approximated by the ones of the unthiolated alkane.}
\label{SI_Table_SolvParams}
\end{table}

\begin{figure}[H]
    \centering
    \includegraphics[width=0.8\linewidth]{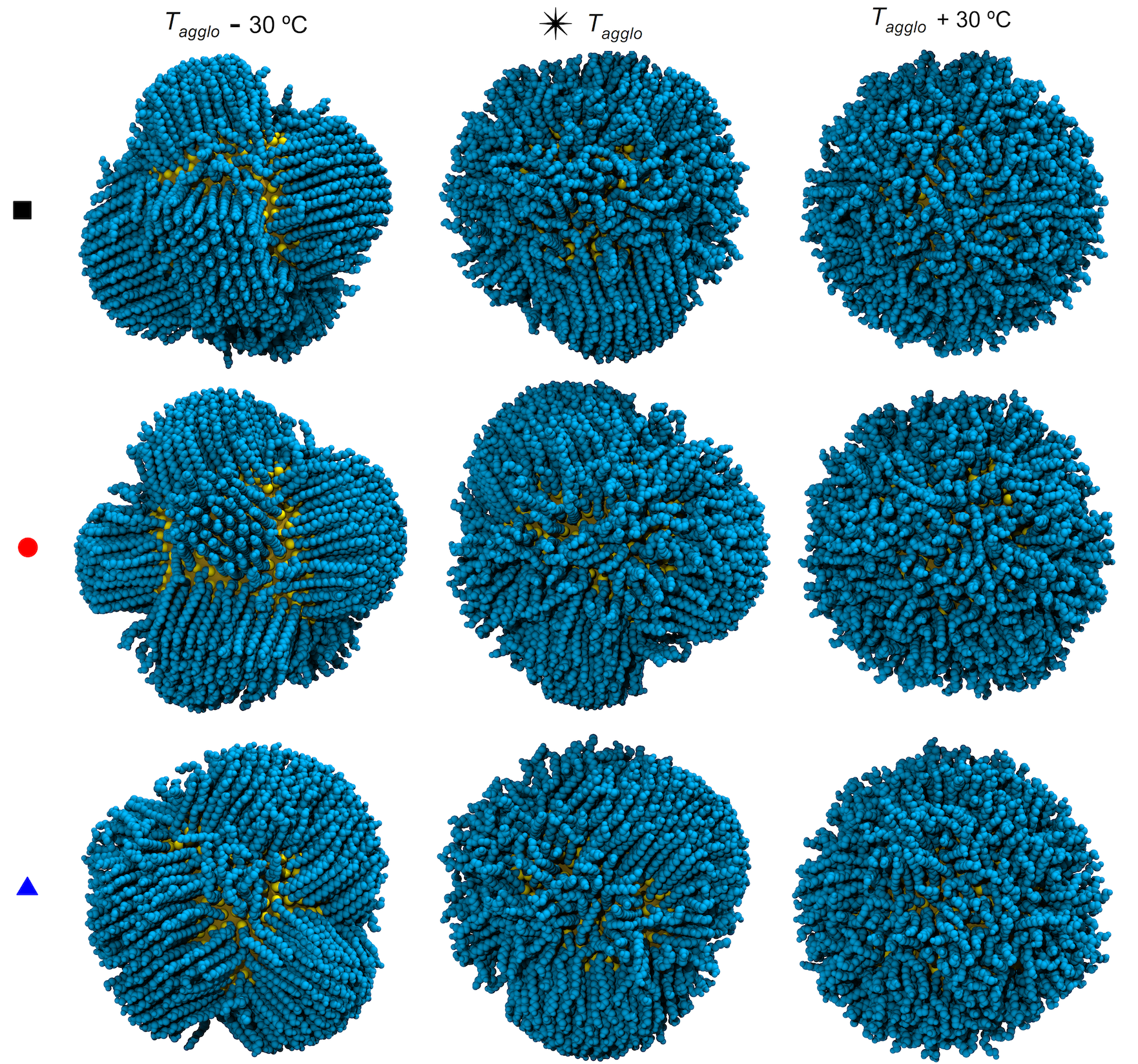}
    \caption{Simulation snapshots of \SI{5.8}{\nano\meter} \ce{CdSe-SC_{18}} particles at the experimental \(T_{agglo}\) $\pm$ \SI{30}{\celsius} in hexane (square), decane (circle), and hexadecane (triangle). Solvent molecules have been hidden for clarity.} 
    \label{SI_CdSe_snapshots}
    \vspace{0.5em}
\end{figure}

\begin{figure}[H]
    \centering
    \includegraphics[width=0.4\linewidth]{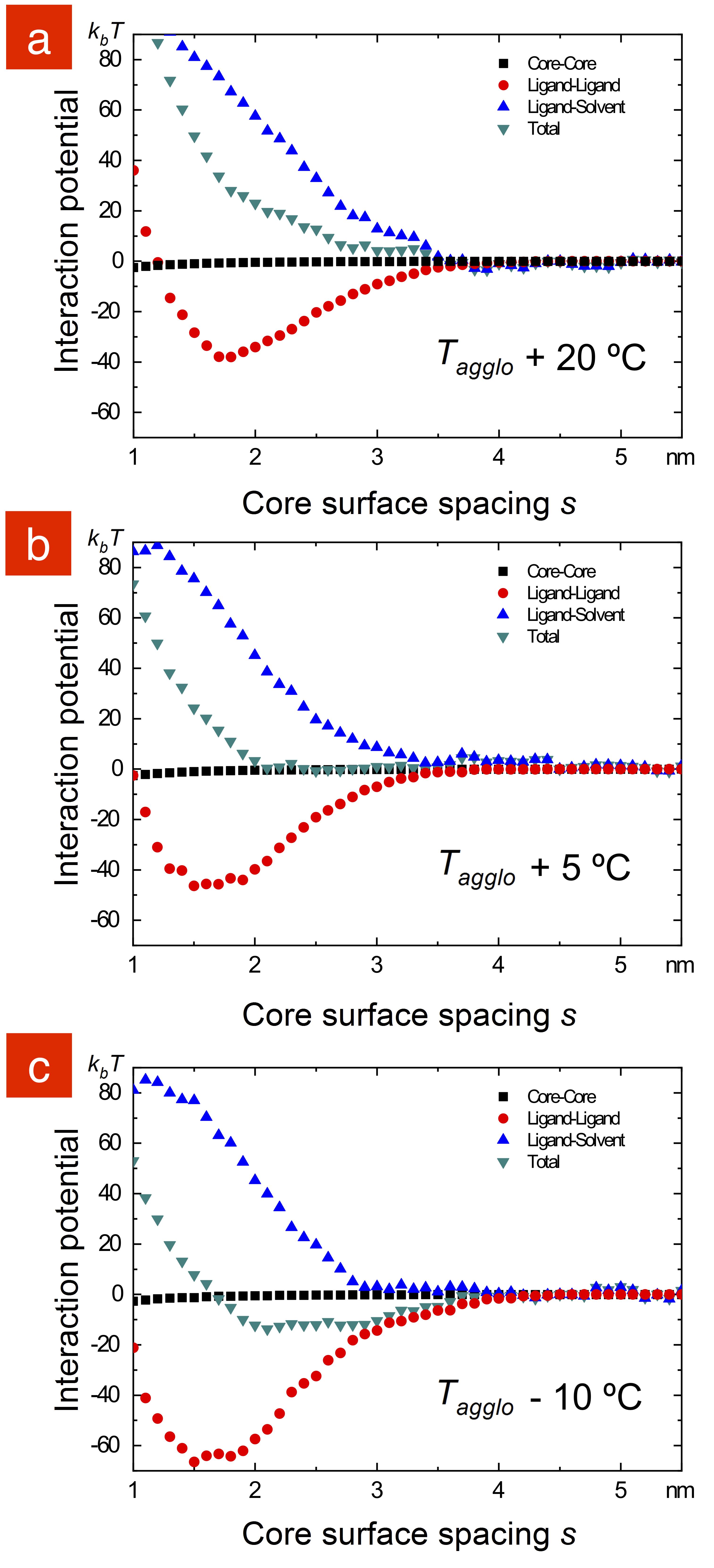}
    \caption{Components of the Potential of Mean Force (PMF) obtained from constrained molecular dynamics simulations for pairs of \SI{4}{\nano\meter} \ce{Au-SC_{16}} particles in hexane at temperatures around the experimental \(T_{agglo}\). The total potential is green, the ligand-ligand contribution is red, the ligand-solvent contribution is blue, and the core-core contribution is black.} 
    \label{SI_PMFComponents}
    \vspace{0.5em}
\end{figure}

Figure \ref{SI_CdSe_snapshots} shows snapshots of our MD simulations of \ce{CdSe}NPs at temperatures around the experimental agglomeration temperature \(T_{agglo}\) in alkane solvents with different chain lengths. Similarly to what is observed for \ce{Au}NPs, the ligands go through a disorder-order transition when the temperature is decreased. This is responsible for changing the overall interaction between the ligand shells, switching the total interaction potential between nanoparticles from repulsive to attractive, as can be seen in Figure \ref{SI_PMFComponents}.

\begin{figure}[H]
    \centering
    \includegraphics[width=0.8\linewidth]{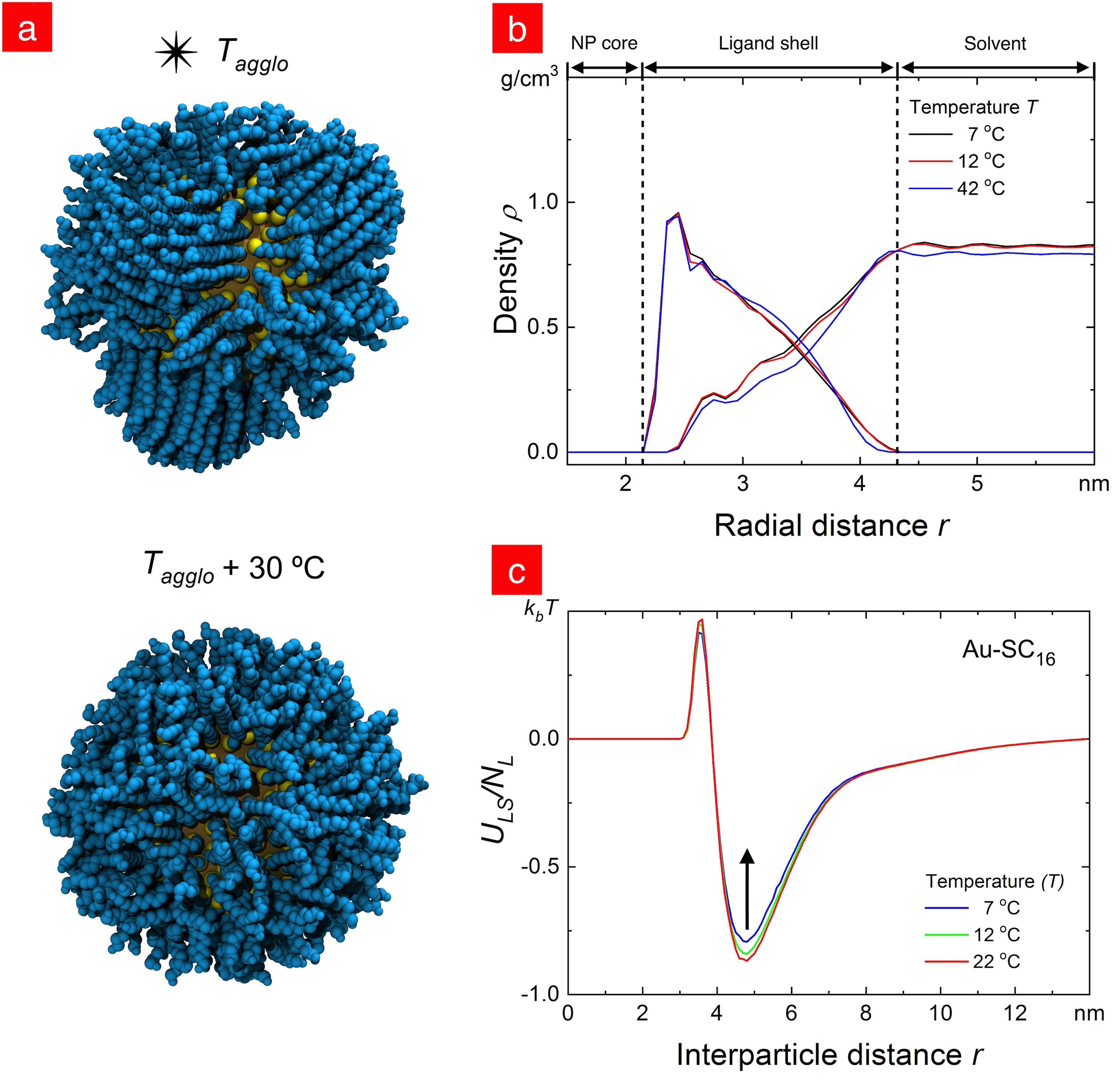}
    \caption{\SI{4}{\nano\meter} \ce{Au-SC_{16}} particles in cyclohexane: (a) Simulation snapshots at the experimental \(T_{agglo}\) and \(T_{agglo}\) $+$ \SI{30}{\celsius}. Solvent molecules have been hidden for clarity. (b) Radial density distributions for the ligand and solvent molecules as a function of the distance \(r\) from the center of the nanoparticle core at (red) and above (blue) \(T_{agglo}\). (c) Contribution to the total ligand-solvent interaction energy as a function of distance between pairs of interacting \ce{CH_{x}} groups. Energies are normalized by the number of ligand molecules on the nanoparticle \(N_L\).}
    \label{SI_Cyclohexane}
    \vspace{0.5em}
\end{figure}

Figure \ref{SI_Cyclohexane} shows snapshots and radial distribution functions for \SI{4}{\nano\meter} \ce{Au-SC_{16}} particles in cyclohexane. The ligands order when the temperature is decreased in a similar way to when the particles are dispersed in linear alkanes. 

\begin{figure}[H]
    \centering
    \includegraphics[width=\linewidth]{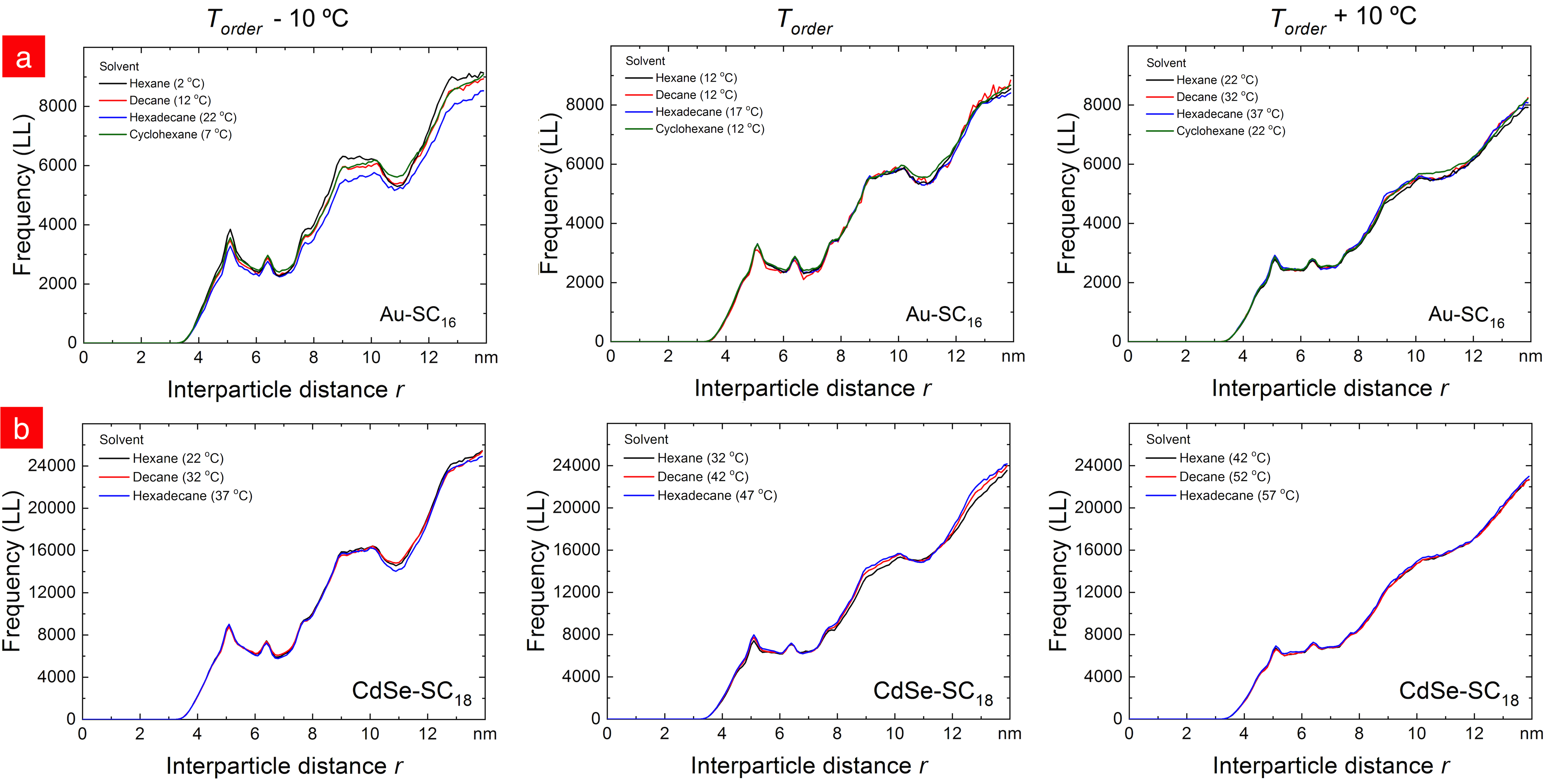}
    \caption{The number of ligand-ligand \ce{CH_{x}-CH_{x}} interactions as a function of the distance between them for (a) \SI{4}{\nano\meter} \ce{Au-SC_{16}} and (b) \SI{5.8}{\nano\meter} \ce{CdSe-SC_{18}} particles shows that the ligands order similarly in all solvents for the same temperature relative to the ordering transition.}
    \label{SI_fll_frequency}
    \vspace{0.5em}
\end{figure}

\begin{figure}[H]
    \centering
    \includegraphics[width=\linewidth]{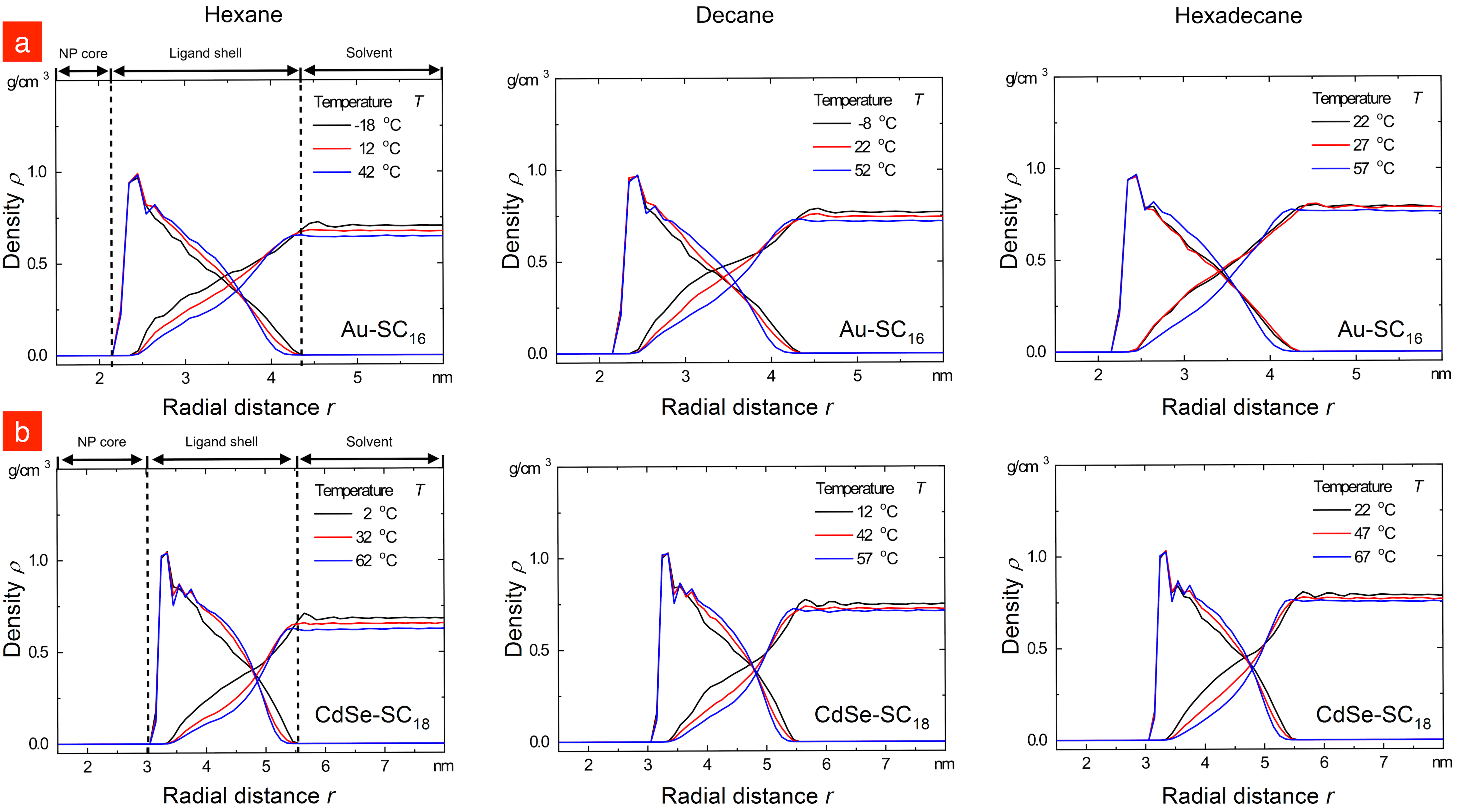}
    \caption{Radial density distributions for the ligand and solvent molecules for \ce{Au} and \ce{CdSe} NP in different alkane solvents and temperatures, plotted as a function of the distance \(r\) from the center of the nanoparticle core. The red lines indicate the density profile at the ligand ordering temperature \(T_{order}\). In all cases, the radially averaged solvent density within the ligand shell increases upon cooling.} 
    \label{SI_Densities}
    \vspace{0.5em}
\end{figure}

Figure \ref{SI_fll_frequency} shows the total number of pair interactions between ligand \ce{CH_{x}} groups as a function of the separation between them, the solvent and the temperature. The temperature at which the ligands order (\(T_{order}\)) differs between solvents, but the average structure of the ligand shell is virtually the same in all solvents when the temperature is expressed relative to \(T_{order}\). Radial density distributions for the ligand and solvent molecules surrounding the particles are shown in Figure \ref{SI_Densities}.

\begin{figure}[H]
    \centering
    \includegraphics[width=\linewidth]{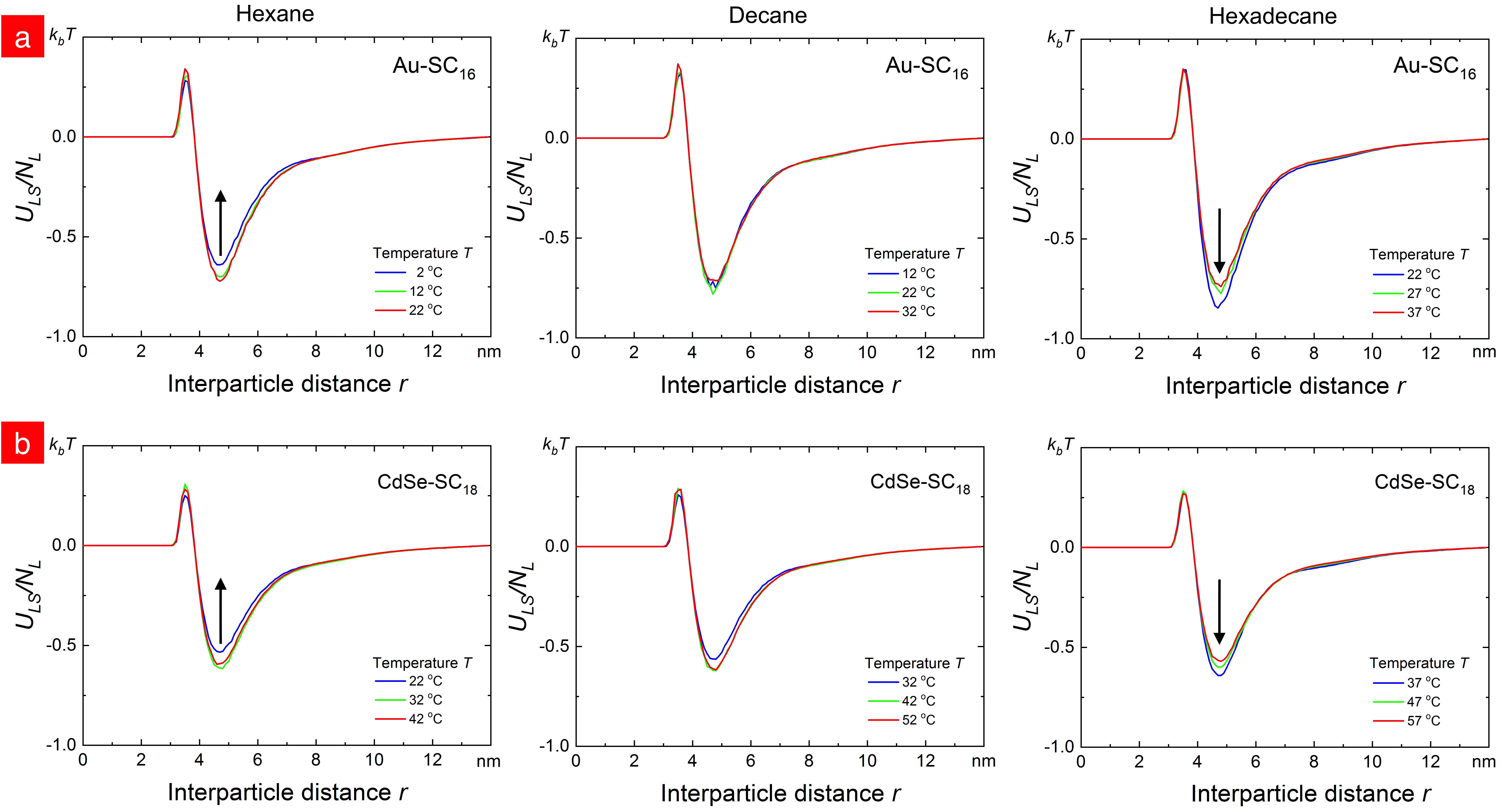}
    \caption{Contribution to the ligand-solvent interaction energy as a function of distance between pairs of interacting \ce{CH_{x}} groups for (a) \SI{4}{\nano\meter} \ce{Au-SC_{16}} and (b) \SI{5.8}{\nano\meter} \ce{CdSe-SC_{18}} particles in different solvents. For longer alkanes, there is an increase in the number of ligand and solvent atoms that are close to one another upon cooling. All energies are normalized by the number of ligand molecules on the nanoparticle \(N_L\).} 
    \label{SI_fls_distribution}
    \vspace{0.5em}
\end{figure}

Figure \ref{SI_fls_distribution} shows the contribution to the ligand-solvent interaction energy as a function of the distance between \ce{CH_{x}} groups. Energies are normalized by the number of ligands on the nanoparticle: 280 ligands for \ce{Au}NP, and 580 ligands for \ce{CdSe}NP.

\begin{figure}[H]
    \centering
    \includegraphics[width=0.8\linewidth]{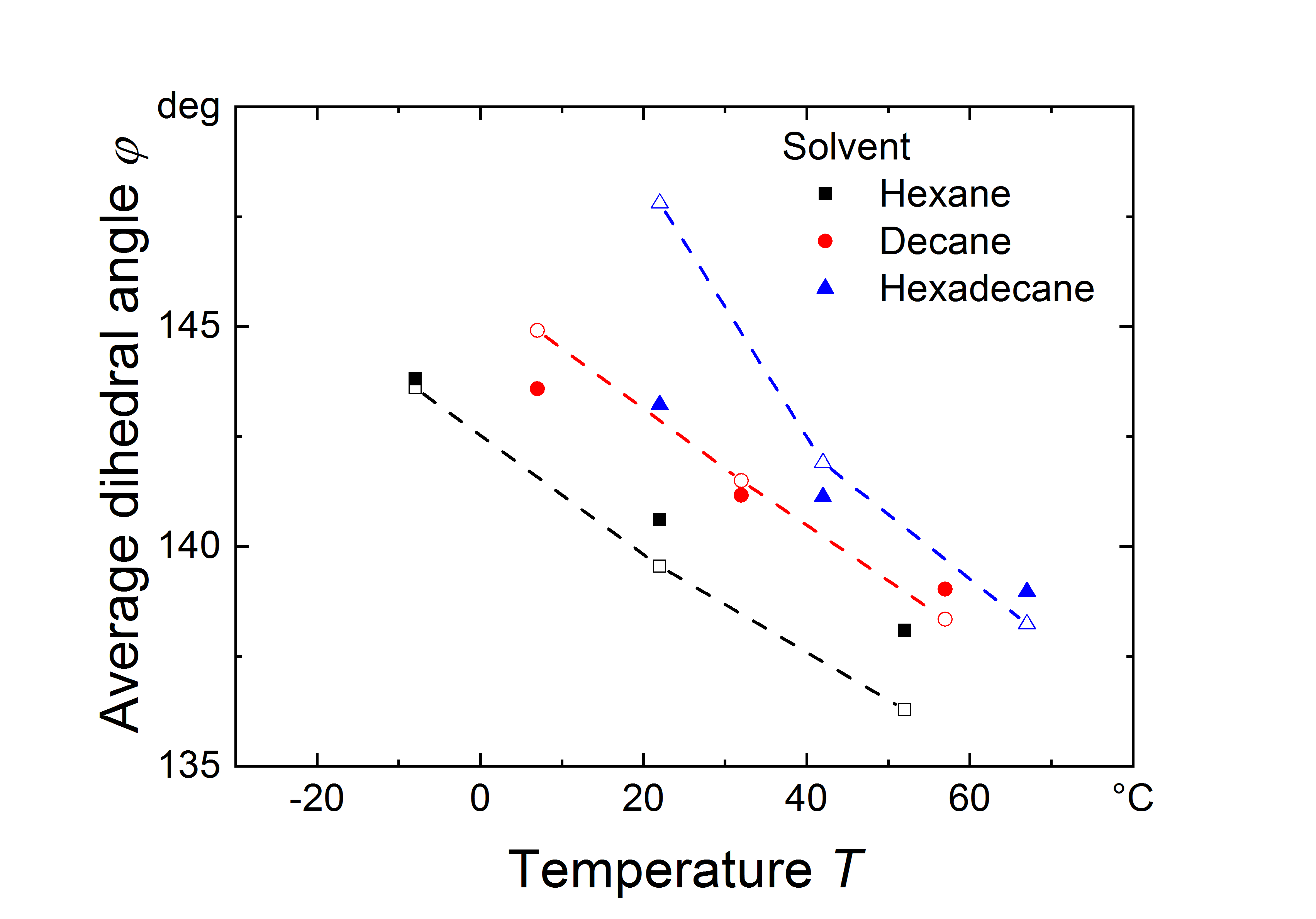}
    \caption{Average dihedral angle of solvent molecules that are in the bulk region (closed symbols) and within the 1-octadecanethiol ligand shell (open symbols) covering \SI{5.8}{\nano\meter} \ce{CdSe}NP.} 
    \label{SI_CdSeDihedral}
    \vspace{0.5em}
\end{figure}

The average dihedral angle of the solvent molecules (Figure \ref{SI_CdSeDihedral} for \ce{CdSe}NPs) was calculated separately for two regions of the simulation cell: (i) within the spherical ligand shell, \textit{i.e.} the region where the radially averaged ligand density is nonzero; and (ii) a region sufficiently far from the nanoparticle that the solvent molecules behave as bulk solvent. 

\begin{table}[!htb]
\caption{Parameters used for analytical calculations of the entropies of mixing using equations 1 and 2.}
\label{tab:mixing_parameters}
\begin{tabular}{ll}
\hline
\multicolumn{1}{c}{parameter}                           & \multicolumn{1}{c}{value} \\ \hline \hline
ligand surface coverage                        & \SI[mode=text]{5.5}{ligands nm^{-2}} \\
ligand length \(L\) for hexadecanethiol                 & \SI{2.28}{\nano\meter} \\
ligand length \(L\) for octadecanethiol                 & \SI{2.54}{\nano\meter} \\
volume ligand molecule \(\nu_L\) for hexadecanethiol   & \SI{0.550}{\nano\meter^3} \\
volume ligand molecule \(\nu_L\) for octadecanethiol   & \SI{0.616}{\nano\meter^3} \\
volume solvent molecule \(\nu_S\) for hexane           & \SI{0.215}{\nano\meter^3} \\
volume solvent molecule \(\nu_S\) for decane           & \SI{0.324}{\nano\meter^3} \\
\multicolumn{1}{l}{volume solvent molecule \(\nu_S\) for hexadecane} & \multicolumn{1}{l}{\SI{0.487}{\nano\meter^3}} \\ \hline \hline
\end{tabular}
\end{table}

\bibliography{bib}